\documentclass[manuscript=article,layout=onecolumn]{achemso} 
\usepackage{graphicx}
\usepackage{amssymb}

\newcommand{\onlinecite}[1]{\hspace{-1 ex} \nocite{#1}\citenum{#1}}

\newcommand{\pvn}{P_v(N)}
\newcommand{\pvnt}{P_v(\tilde{N})}

\newcommand{\fvnt}{F_v(\tilde{N})}

\newcommand{\Nt}{\tilde{N_v}}
\newcommand{\avgNt}{\langle \tilde{N_v} \rangle}
\newcommand{\varNt}{\langle \delta \tilde N_v^2 \rangle}

\newcommand{\deltaNt}{\delta(\tilde N_v- \tilde N)}

\newcommand{\avgN}{\langle N \rangle}

\title{Sparse Sampling of Water Density Fluctuations in Interfacial Environments}
\author{Erte Xi}
\author{Richard C. Remsing}
\author{Amish J. Patel}
\affiliation[University of Pennsylvania]
{Department of Chemical and Biomolecular Engineering, University of Pennsylvania, Philadelphia, PA 19104, USA}
\email{amish.patel@seas.upenn.edu}
\begin{document}
\begin{abstract}
The free energetics of water density fluctuations near a surface, and the rare low-density fluctuations in particular, serve as reliable indicators of surface hydrophobicity; the easier it is to displace the interfacial waters, the more hydrophobic the underlying surface.
However, characterizing the free energetics of such rare fluctuations requires computationally expensive, non-Boltzmann sampling methods like umbrella sampling.
This inherent computational expense associated with umbrella sampling makes it challenging to investigate the role of polarizability or electronic structure effects in influencing interfacial fluctuations.
Importantly, it also limits the size of the volume, which can be used to probe interfacial fluctuations.
The latter can be particularly important in characterizing the hydrophobicity of large surfaces with molecular-level heterogeneities, such as those presented by proteins.
To overcome these challenges, here we present a method for the sparse sampling of water density fluctuations, which is roughly two orders of magnitude more efficient than umbrella sampling.
We employ thermodynamic integration to estimate the free energy differences between biased ensembles, thereby circumventing the umbrella sampling requirement of overlap between adjacent biased distributions.
Further, a judicious choice of the biasing potential allows such free energy differences to be estimated using short simulations, so that the free energetics of water density fluctuations are obtained using only a few, short simulations.
Leveraging the efficiency of the method, we characterize water density fluctuations in the entire hydration shell of the protein, ubiquitin; a large volume containing an average of more than six hundred waters.
\end{abstract}


\section{Introduction}
%
An understanding of density fluctuations in bulk water and at interfaces, has played a central role in the description of hydrophobic effects~\cite{FHS:1973,Pratt:JCP:1977,Hummer:PNAS:1996,ashbaugh_SPT,LCW,HGC,Huang:JPCB:2002,Chandler:Nature:2005,mittal_pnas08,garde09prl,Godawat:PNAS:2009,Acharya:Faraday:2010,Patel:JPCB:2010,Patel:PNAS:2011,LLCW,Jamadagni:ARCB:2011,Chandler:2012,Patel:JPCB:2012,Remsing:2013,Patel:JPCB:2014}, which drive biomolecular~\cite{dill_rev02,Dobson:2003,Levy:2006aa,shea08,ball08,Thirumalai:2012} and other aqueous assemblies~\cite{Tanford1973,Whitesides:2002aa,Rabani:2003,MaibaumDinnerChandler2004,Morrone:2012}.
For example, the fact that $P_v(N)$, the probability of observing $N$ water molecules in a small observation volume ($v\lesssim 1$~nm$^3$), 
obeys Gaussian statistics~\cite{Hummer:PNAS:1996, Garde:PRL:1996}, has provided molecular-level insights into the the pressure-induced denaturation of proteins~\cite{tW_rev,pratt_rev,Hummer:PNAS:1998},  as well as the convergence of protein unfolding entropies at a particular temperature~\cite{Garde:PRL:1996,Remsing:2013}. 
Similarly, the Lum--Chandler--Weeks (LCW) theory~\cite{LCW} prediction that $P_v(N)$ should develop fat low-$N$ tails for large volumes ($v\gtrsim 1$~nm$^3$) in bulk water, and near hydrophobic surfaces~\cite{LCW,HuangChandlerPRE,LLCW}, and its verification by simulations~\cite{Patel:JPCB:2010,Patel:PNAS:2011,Patel:JPCB:2012,Remsing:JCP:2015}, has clarified that water near hydrophobic surfaces is sensitive to perturbations~\cite{Patel:JPCB:2012,Levy:1992,berne04,berne05_melittin,chou_dewet,chou05,garde_rev}, and led to the insight that extended hydrophobic surfaces could generically serve as catalysts for the assembly and disassembly of small hydrophobic solutes~\cite{Patel:PNAS:2011,Vembanur:2013}.
%

%
Importantly, 
the fact that low-$N$ fluctuations are enhanced near hydrophobic surfaces, also makes them a suitable metric for quantifying the hydrophobicity of a surface, or the strength of its interactions with water~\cite{Patel:JPCB:2010,Jamadagni:ARCB:2011,Patel:JPCB:2012,Patel:JPCB:2014}. 
Fluctuations are particularly useful in characterizing the hydrophobicity of complex surfaces with molecular-level heterogeneities, wherein conventional macroscopic measures, such as the water droplet contact angle, break down~\cite{Granick:Science:2008}.
Indeed, for the rugged, heterogenous  surfaces of proteins, hydrophobicity depends 
not only on the chemistry of the underlying residues~\cite{Siebert:Biochem:2002,Acharya:Faraday:2010,Patel:JPCB:2012,Patel:JPCB:2014}, but also on the particular topography~\cite{Giovambattista:PNAS:2008,Mittal:Faraday:2010,Luzar:Faraday:2010} and chemical pattern~\cite{Giovambattista:JPCC:2007,Berne:JPCC:2009,Striolo:Langmuir:2009,Acharya:Faraday:2010,Luzar:PNAS:2011} presented by the protein, and can depend non-trivially on the specific combination of the two~\cite{berne04,berne05_melittin,Giovambattista:PNAS:2009, Surblys:JCP:2011, Rotenberg:JACS:2011}.
Fluctuations have previously been used to characterize the hydrophobicity of such complex surfaces using 
small (e.g., methane-sized) $v$~\cite{Siebert:Biochem:2002,Acharya:Faraday:2010}, 
wherein the fluctuations of interest can be estimated using equilibrium molecular simulations~\cite{Widom:JCP:1963,Hummer:PNAS:1996}.
However, the likelihood of low-$N$ fluctuations decreases roughly exponentially with the size of $v$, and estimating $\pvn$ for larger $v$ requires computationally expensive biased sampling techniques, such as umbrella sampling~\cite{DCbook,Patel:JPCB:2010,Patel:JSP:2011}. 
In particular, the larger the volume of interest, the larger the number of biased simulations required.
As a result, using large volumes, such as the hydration shells of entire proteins, to characterize surface hydrophobicity would be very expensive. 
Further, using umbrella sampling in conjunction with more computationally expensive treatments, which incorporate the polarizability of interfacial waters or electronic structure effects~\cite{warren2008electrostatic,remsing2014role}, to characterize density fluctuations in large volumes, would also be prohibitively expensive.
%

%
Here we present a method that enables estimation of $\pvn$ at a number of sparsely separated $N$-values, albeit with a roughly two orders of magnitude increase in computational efficiency, as compared with conventional techniques, such as umbrella sampling or free energy perturbation.
To facilitate sparse sampling, we employ thermodynamic integration to estimate free energy differences between biased ensembles, thereby circumventing the umbrella sampling requirement that adjacent biased distributions overlap.
Building upon recent work by Patel and Garde~\cite{Patel:JPCB:2014}, we employ a linear biasing potential, which enables efficient estimation of free energy differences between biased ensembles using averages that converge rapidly and require only short simulations.
In the following section, we first derive the central equations underlying our method, followed by details pertaining to the systems studied and the simulations employed. 
In the subsequent section, we demonstrate how the method works using a small volume in bulk water. 
We then apply the method to characterize the hydrophobicity of interfaces; first, uniform self-assembled monolayer surfaces, and then the entire hydration shell of the protein, ubiquitin. 
We then discuss the underpinnings of the method's efficiency and its limitations, and conclude with a discussion of scenarios where the method may find broader applicability.
%

\section{Theory and Methods}
\subsection{Derivation of the Central Equations}
%
Consider an observation volume, $v$, described by its size, shape, and location.
We are interested in characterizing the statistics of water density fluctuations in $v$, as quantified by the probability, $\pvn\equiv\langle \delta(N_v - N) \rangle_0$, of observing $N$ water molecules in $v$.
Here, $\langle \dots \rangle_0$ corresponds to an average in the presence of a generalized Hamiltonian, $\mathcal{H}_0$, and the operator, $N_v(\{\mathbf r_i\})$, depends on the positions, $\{\mathbf r_i\}$, of all the water oxygens.
To circumvent issues related to the sampling of the discrete operator, $N_v(\{\mathbf r_i\})$, using molecular dynamics (MD) simulations, here we will focus on quantifying the closely related probability, $\pvnt\equiv\langle \delta(\Nt-\tilde N) \rangle_0$, of observing $\tilde N$ coarse-grained water molecules in $v$.
The operator, $\tilde N_v(\{\mathbf r_i\})$, is chosen to be strongly correlated with the (discrete) operator, $N_v(\{\mathbf r_i\})$, but is a continuous function of $\{\mathbf r_i\}$, so that $\Nt$-dependent biasing potentials do not give rise to impulsive forces in MD simulations.
The functional forms that we employ for $\tilde N_v(\{\mathbf r_i\})$, can be found in ref.~\onlinecite{Patel:JSP:2011}.
Importantly, as shown in the Supporting Information, in addition to enabling sparse sampling of $\pvnt$, the method we present below, can also be used to obtain $\pvn$.
%

To estimate $\pvnt$ over the entire range of $\tilde N$-values, including those that are highly improbable, standard approaches such as umbrella sampling, prescribe the application of a biasing potential that is a function of $\Nt$~\cite{DCbook,Frenkel_Smit}.
Consider the biasing potential, $U_{\phi} (\tilde N_v)$, parametrized by $\phi$, so that the Hamiltonian of the biased system becomes $\mathcal{H}_{\phi} = \mathcal{H}_0 + U_{\phi} (\tilde N_v)$; let $Q_{\phi}$ be the partition function associated with $\mathcal{H}_{\phi}$.
The unbiased probability, $P_v(\tilde N)$, of observing $\tilde N$ coarse-grained waters in $v$ can be readily related to the corresponding probability in the biased ensemble, $P_v^{\phi}(\tilde N)$, as
\begin{equation}
P_v(\tilde N) =  P_v^{\phi}(\tilde N) e^{\beta U_{\phi}(\tilde N)} \left(\frac{Q_{\phi}}{Q_0}\right),
\label{eq:us}
\end{equation}
where $\beta\equiv1/k_{\rm B}T$, with $k_{\rm B}$ being the Boltzmann constant and $T$ being the temperature.
A derivation for Equation~\ref{eq:us} is included in the Supporting Information.
Taking the logarithm of both sides and defining the free energy, $F_v(\tilde N)$, of observing $\tilde N$ coarse-grained waters in $v$, through $\beta F_v(\tilde N)\equiv -\ln P_v(\tilde N)$, we get the standard result that serves as a starting point for analyzing umbrella sampling results~\cite{DCbook,Frenkel_Smit},
\begin{equation}
F_v(\tilde N) =  F_v^{\phi} (\tilde N) - U_{\phi} (\tilde N) + F_{\phi}.
\label{eq:us2}
\end{equation}
Here, $\beta F_v^{\phi} (\tilde N)$ is similarly defined as $-\ln P_v^{\phi}(\tilde N)$, and $\beta F_{\phi} \equiv -\ln \left(\frac{Q_{\phi}}{Q_0}\right)$ is the free energy difference between the biased and unbiased ensembles.
The first term in Equation~\ref{eq:us2} can be readily estimated from biased simulations, and the second term is  trivial to evaluate, because the functional form of the biasing potential is known.
In contrast, the third term, $F_{\phi}$, is not known a priori; umbrella sampling relies on overlap in adjacent biased distributions to estimate $F_{\phi}$, usually with the help of algorithms such as WHAM or MBAR~\cite{roux_wham,wham2,MBAR}.
Such a prescription provides accurate estimates of $F_v(\tilde N)$ over the entire range of sampled $\Nt$-values.

%
Nevertheless, obtaining a continuous $F_v(\tilde N)$ distribution is not always necessary; instead, a knowledge of its functional form at $\tilde N$-values that are well separated can often be sufficient.
However, the umbrella sampling requirement of overlap between adjacent biased distributions is incommensurate with the spirit of such sparse sampling. %
Importantly, the overlap requirement also necessitates a large number of biased simulations, and can be computationally expensive, in particular for large $v$.
The proposed method aims to circumvent this overlap requirement by estimating $F_{\phi}$ using thermodynamic integration.
We will demonstrate that such an approach can be both accurate and highly efficient for a particular choice of the biasing potential, $U_{\phi} (\tilde N_v)$.
In particular, we build upon recent work by Patel and Garde~\cite{Patel:JPCB:2014}, which employed a linear biasing potential to estimate the free energy of emptying $v$, as quantified by $F_v(\tilde N=0)$, with a roughly 2 orders of magnitude decrease in computational effort.
Here we show how the entire $P_v(\tilde N)$ distribution can be sparsely sampled with a similar increase in computational efficiency, through the use of such a linear potential, $U_{\phi} (\tilde N_v)=\phi \tilde N_v$, whose strength is determined by the parameter, $\phi$.
In the presence of the linear potential, $U_\phi$, Equation~\ref{eq:us2} becomes
\begin{equation}
F_v(\tilde N) = F_v^{\phi} (\tilde N) -\phi \tilde N + F_{\phi}.
\label{eq:us-phi}
\end{equation}
As discussed above, the first two terms of the above equation are readily determined. 
Here we determine $F_{\phi}$ using thermodynamic integration, that is, by integrating $\partial F_\phi/\partial\phi = \langle \tilde N_v \rangle_{\phi}$ as:
\begin{equation}
F_{\phi} = \int_0^\phi \langle \tilde N_v \rangle_{\phi'} \mathrm{d}\phi'.
\label{eq:fphi}
\end{equation}
The integration is performed over a range of $\phi$-values from zero (i.e., the unbiased system) to the particular $\phi$-ensemble of interest.
Increasing $\phi$ penalizes the presence of water molecules in $v$; indeed, because $\partial^2 F_\phi/\partial\phi^2 = \partial\langle \tilde N_v \rangle_{\phi}/\partial\phi = -\beta\langle \delta \tilde N_v^2 \rangle_{\phi}<0$,  $\avgNt_\phi$ decreases monotonically as $\phi$ is increased.  
To illustrate how the method works, consider employing Equation~\ref{eq:us-phi} for $\tilde N=\avgNt_\phi$,
\begin{equation}
F_v \bigg( \avgNt_\phi \bigg) = F_v^{\phi} \bigg( \avgNt_\phi \bigg) - \phi \avgNt_\phi + \int_0^\phi \langle \tilde N_v \rangle_{\phi'} \mathrm{d}\phi'.
\label{eq:central-phi2} 
\end{equation}
Thus, from estimates of the average values, $\avgNt_{\phi_k}$, obtained from a series of biased simulations with potentials of strength, $\{\phi_k\}$, the corresponding estimates of $F_v(\tilde N = \avgNt_{\phi_k})$ can be readily obtained at $k$ distinct $\tilde N$-values.
In addition to $F_v(\tilde N = \avgNt_{\phi_k})$, this method can also be used to obtain estimates of $F_v(\tilde N)$ in the vicinity of $\tilde N=\avgNt_{\phi_k}$; in particular, using Equations~\ref{eq:us-phi} and~\ref{eq:fphi}, $F_v(\tilde N)$ can be estimated for each $\tilde N$-value sampled by any of the biased simulations.

%
We note that while we have chosen to focus on the fluctuations in $\Nt$, the method presented here can readily be generalized to chararcterize fluctuations in any order parameter.
Further, as mentioned above and shown in the Supporting Information, this approach can also be used to indirectly sample discrete order parameters; in particular, $\pvn$, the probability of observing $N$ water molecules in $v$, can be readily estimated.
Finally, we note that while there are certain similarities between our method and the Umbrella Integration (UI) method, proposed by K{\"a}stner and Thiel~\cite{Kastner:JCP:2005}, there are important fundamental and practical differences between the two methods; a detailed discussion of these issues is included in the Supplementary Information.

\subsection{Simulation Details}
We illustrate the application of Equation~\ref{eq:central-phi2} to compute the $F_v(\tilde N)$ for 3 different classes of systems:
(1) a small spherical volume of radius $R_v=0.5$~nm in bulk water;
(2) a cylindrical disk of width, $w=0.3$~nm and radius, $R_v=2$~nm, in the hydration shell of self-assembled monolayers (SAMs) of alkyl chains with two different head groups: -OH and -CH$_3$, which correspond to hydrophilic and hydrophobic SAM-water interfaces respectively; and
(3) the entire hydration shell of the protein, ubiquitin (PDB: 1UBQ)~\cite{ubiquitin}, as defined by the union of spherical sub-volumes of radii, $r_v=0.6$~nm, centered on each of the protein heavy atoms. 
%

To study each of these systems, we use all-atom MD simulations using the GROMACS package~\cite{gmx4ref}, suitably modified to incorporate the biasing potentials of interest. 
We perform biased simulations at different $\phi$-values, and use our central Equation~\ref{eq:central-phi2} to estimate $F_v(\tilde N)$ at sparsely separated $\tilde N$-values. 
In all cases, a simulation box with periodic boundary conditions was employed, and the leap frog algorithm~\cite{Frenkel_Smit} with a 2 fs time-step was used to integrate the equations of motion. 
For the bulk water and SAM systems, the SPC/E model of water was used~\cite{SPCE}, whereas the protein was hydrated using TIP4P water~\cite{TIP4P}.
The short-range Lennard-Jones and electrostatic interactions were truncated at 1~nm, whereas the long range electrostatic interactions were treated using the Particle Mesh Ewald algorithm~\cite{PME}. 
Bonds involving hydrogen atoms were constrained; the SHAKE algorithm~\cite{SHAKE} was employed for the bulk water system, and the LINCS algorithm~\cite{LINCS} was used for the SAM and protein systems.
In all cases, a constant temperature of 300~K was maintained by using the canonical velocity-rescaling thermostat~\cite{V-Rescale}. 
For the bulk water system, the pressure was additionally maintained at 1~bar using the Parrinello-Rahman barostat~\cite{Parrinello-Rahman}. 
While the SAM and protein systems are simulated in the canonical ensemble, a buffering water-vapor interface was employed to ensure that the system remains at coexistence pressure~\cite{Miller:PNAS:2007,Patel:JPCB:2010,Patel:JSP:2011}. 
Additional system specific details including those pertaining to the placement of observation volumes are given below. 
\\ {\it Bulk Water}:
The spherical observation volume is situated at the center of a cubic simulation box with a side of length 6~nm.
\\ {\it SAM Surfaces}:
The simulation setup is similar to that used in ref.~\onlinecite{Patel:PNAS:2011}. 
The disk-shaped observation volumes are placed adjacent to the SAM surfaces, with their axes perpendicular of the surface.
For the CH$_3$-terminated SAM surface, the observation volume is centered at the first peak of the water density distribution in the direction perpendicular of the SAM surface.  
For the OH-terminated SAM surface, the proximal edge of the observation volume is placed at a location where the averaged density of the SAM heavy atoms drops below 2\% of the bulk water density.
\\ {\it Protein Hydration Shell}:
We used the CHARMM27 forcefield~\cite{CHARMM} to simulate ubiquitin, which was hydrated by roughly 13,000 TIP4P water molecules~\cite{TIP4P}. 
A 3~ns unbiased simulation was first run to equilibrate the hydrated protein. 
To ensure that protein atoms remain in the observation volume, all protein heavy atoms are position restrained harmonically with a relatively soft spring constant of 1000~kJ/nm$^2$ in each dimension.
Each biased simulation was then run for 3 ns.
The first 100 ps is discarded as equilibration in response to the biasing potential, and 
the subsequent 100~ps is used to estimate $\langle N_v \rangle_\phi$, whereas the subsequent 2.9~ns were used to estimate $\langle \delta N_v^2 \rangle_\phi$ and to obtain accurate estimates of $\pvn$ using umbrella sampling in conjunction with WHAM.
To obtain $\pvn$ using umbrella sampling, several additional biased simulations are also needed to ensure overlap between adjacent biased distributions.   
%


\begin{figure}[htbp]
\begin{center}
\includegraphics[width=0.95\textwidth]{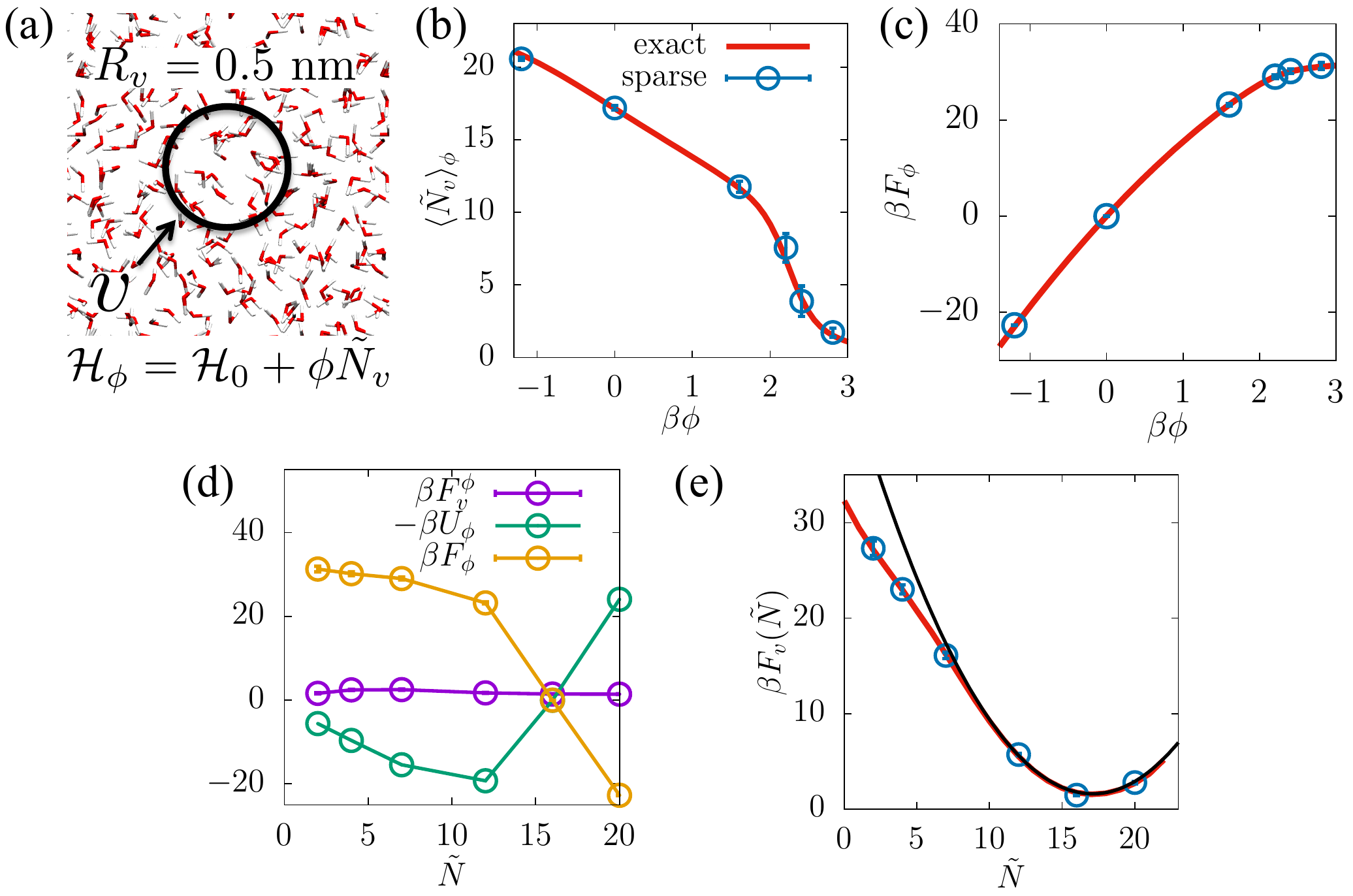}
\caption{
Using the sparse sampling method (Equation~\ref{eq:central-phi2}) to estimate the free energetics of water density fluctuations, $\beta F_v(\tilde N)\equiv-\ln\pvnt$, in a sphere of radius, $R_v=5$~\AA, in bulk water. 
(a) Simulation snapshot highlighting the observation volume, $v$.
(b) The response of the average number of waters, $\avgNt_\phi$ in $v$, to the strength, $\phi$, of a linear biasing potential, $U_\phi=\phi\Nt$, is estimated using biased simulations.
(c) The free energy difference between the biased and unbiased ensembles, $F_\phi$, is estimated by integrating the response of $\avgNt_\phi$ to $\phi$; see Equation~\ref{eq:fphi}.
(d) The three components of $\beta F_v(\tilde N)$ according to Equation~\ref{eq:central-phi2}.
(e) The sparsely sampled water density fluctuations obtained using the $\avgNt_\phi$-values in panel b, agree well with the exact fluctuation spectrum obtained using umbrella sampling.
Comparison with the corresponding Gaussian distribution (black line) highlights the presence of a low-$\tilde N$ fat tail in $\pvnt$.
}
\label{fig:bulk}
\end{center}
\end{figure}

\section{Results and Discussion}

\subsection{Bulk Water}
%
Figure~\ref{fig:bulk} illustrates the sparse sampling method for estimating $F_v(\tilde N)$ using a small spherical volume, $v$, of radius, $R_v=0.5$~nm, in bulk water (Figure~\ref{fig:bulk}a).
$v$ contains roughly $\avgNt_0\approx16$~water molecules on average.
Figure~\ref{fig:bulk}b shows how the average number of waters in $v$, $\avgNt_\phi$, responds to the linear biasing potential, $U_\phi=\phi \Nt$.
$\avgNt_\phi$ decreases monotonically as the strength of the unfavorable potential, $\phi$, is increased; the decrease is linear in the vicinity of $\phi=0$, but becomes more pronounced at larger $\phi$-values.
Integrating this response according to Equation~\ref{eq:fphi}, enables us to estimate the free energies, $F_\phi$, of the biased ensembles relative to that of the unbiased ensemble (Figure~\ref{fig:bulk}c).
Each of the 3 terms that go into the estimation of $F_v(\tilde N)$ using Equation~\ref{eq:us2} can then be readily obtained at $\tilde N=\avgNt_\phi$, and are shown in Figure~\ref{fig:bulk}d.
The biased distribution free energies, $F_v^\phi(\tilde N=\avgNt_\phi)$, are small because typical water number distributions are peaked at their means; consequently, $P_v^\phi(\tilde N)$ is highest for $\tilde N=\avgNt_\phi$, and $\beta F_v^\phi(\tilde N)\equiv-\ln P_v^\phi(\tilde N)$ is correspondingly small at $\tilde N=\avgNt_\phi$.
The sparse sampled $F_v(\tilde N)$ resulting from the addition of the three terms in Figure~\ref{fig:bulk}d, displays a fat low-$\Nt$ tail as shown in Figure~\ref{fig:bulk}e, and is in excellent agreement with the exact $F_v(\tilde N)$ obtained by umbrella sampling.

\begin{figure}
\begin{center}
\includegraphics[width=0.75\textwidth]{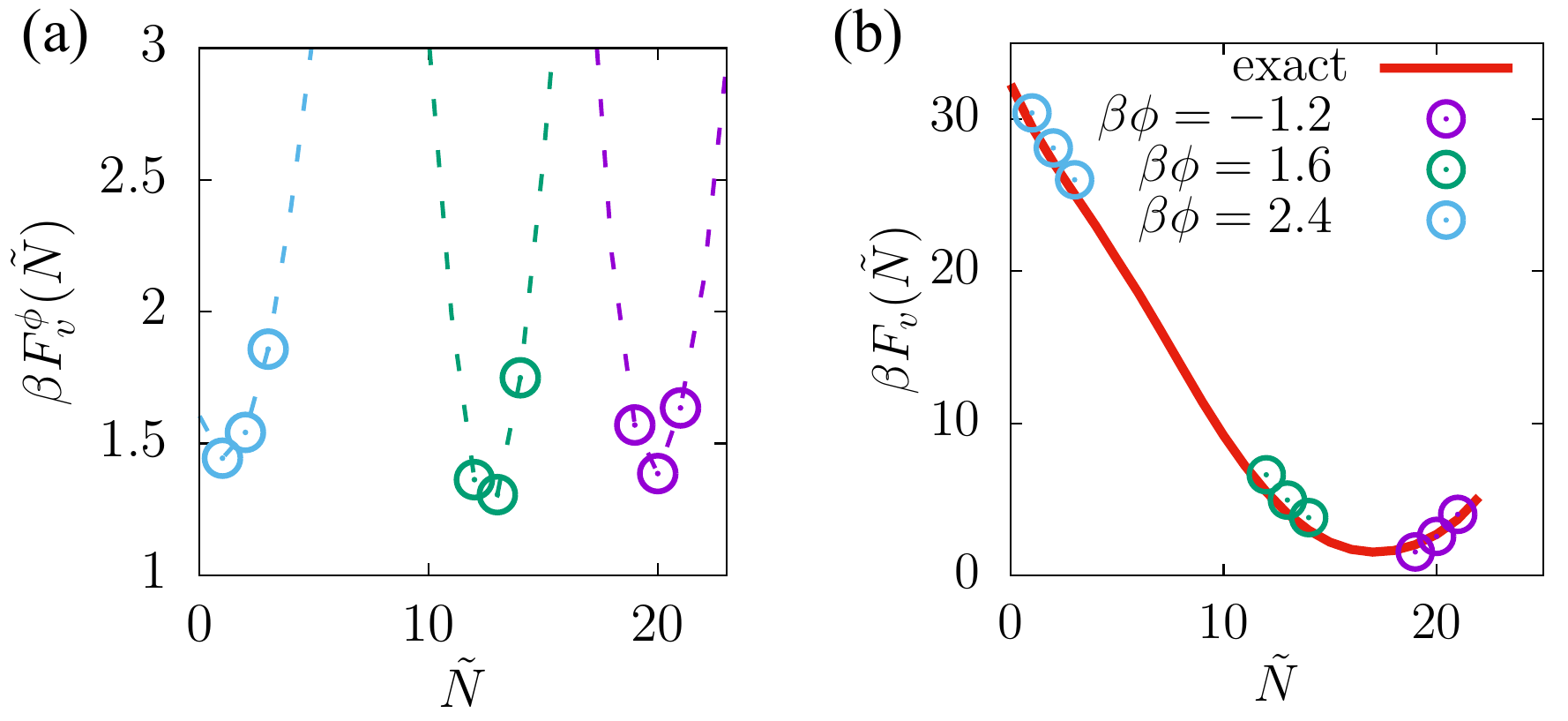}
\caption{
Computing $\fvnt$ for all sampled $\tilde N$-values (in addition to $\tilde N = \avgNt_\phi$).
(a) The free energetics of $\tilde N$-fluctuations obtained from three biased simulations ($\beta \phi = -1.2, 1.6, 2.4$) are shown (dashed lines). Three representative $\tilde N$-values in the vicinity of $\tilde N=\avgNt_\phi$ are highlighted for each biased distribution (symbols).
(b) $\fvnt$ is estimated at each of these $\tilde N$-values using Equations~\ref{eq:us-phi} and~\ref{eq:fphi}, and agrees well with the exact results obtained by umbrella sampling.
}
\label{fig:mult}
\end{center}
\end{figure}

\noindent{\bf Obtaining $\fvnt$ in the vicinity of $\tilde N=\avgNt$ }\\
In addition to estimating $\fvnt$ at $\tilde N=\avgNt_\phi$ using Equation~\ref{eq:central-phi2}, $\fvnt$ can be accurately estimated at other well-sampled $\tilde N$-values using Equations~\ref{eq:us-phi} and~\ref{eq:fphi}. 
In Figure~\ref{fig:mult}, we demonstrate this by calculating $\fvnt$ at three different $\tilde N$-values each, for three of the biased simulations shown in Figure~\ref{fig:bulk}.
Figure~\ref{fig:mult}a shows the biased distributions, $F_v^{\phi} (\tilde N)$, obtained at all the $\tilde N$-values sampled from the biased simulations (dashed lines).
Also highlighted with symbols, are the $F_v^{\phi} (\tilde N)$-values at three well-sampled $\tilde N$-values, chosen in the vicinity of $\tilde N=\avgNt_\phi$.
Using these $F_v^{\phi} (\tilde N)$-values and the $F_\phi$-values shown in Figure~\ref{fig:bulk}c, the sparse sampled $\fvnt$-values obtained from the three biased simulations are shown in Figure~\ref{fig:mult}b, and agree well with the umbrella sampling results.
In principle, $\fvnt$ can thus be estimated at every $\tilde N$-value sampled in a biased simulation.
However, $\tilde N$-values far from the mean, $\avgNt_\phi$, may be insufficiently sampled in practice, leading to significant uncertainties in estimates of $F_v^{\phi} (\tilde N)$, and correspondingly of $F_v(\tilde N)$.
%

\begin{figure}[htbp]
\begin{center}
\includegraphics[width=0.75\textwidth]{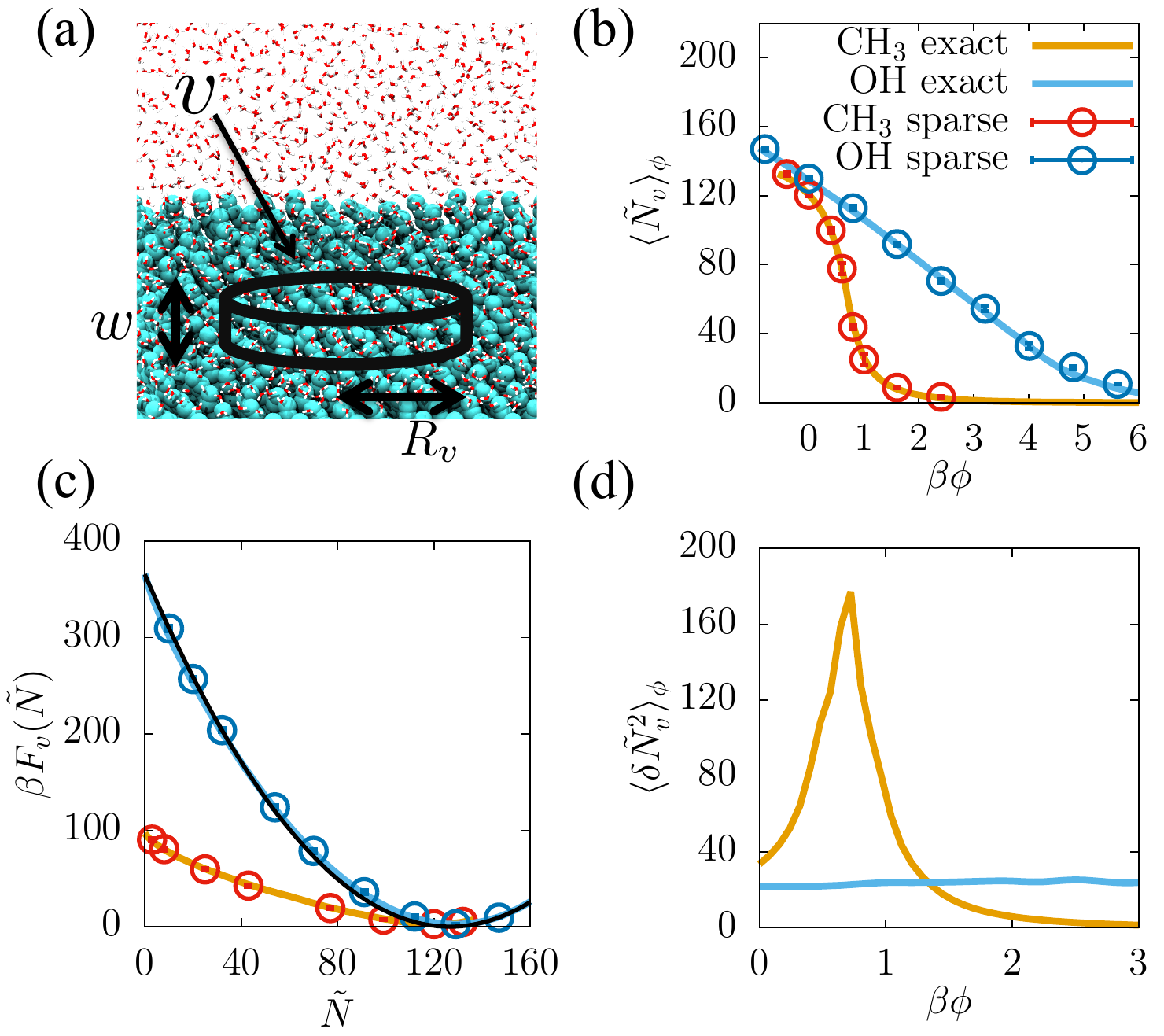}
\caption{
Estimating interfacial water density fluctuations using sparse sampling.
(a) A representative simulation snapshot of the interface between a CH$_3$-terminated self-assembled monolayer (SAM) surface and water.
The alkyl chains (cyan) are shown in spacefill representation, whereas the water molecules (red/white) are shown as sticks. 
The observation volume, $v$ (black lines), is a cylindrical disk of radius $R_v=2$~nm, and thickness $w=0.3$~nm that is situated at the SAM-water interface.
(b) The response of the average number of interfacial waters, $\avgNt_\phi$, to the strength of the linear biasing potential, $\phi$, depends on the hydrophobicity of SAM-water interface.
While the response is roughly linear for the hydrophilic OH-terminated SAM surface, it is sigmoidal for the hydrophobic CH$_3$-terminated SAM surface~\cite{Patel:JPCB:2012}.
(c) The disparate responses arise from a fundamental difference in the statistics of the underlying interfacial density fluctuations; while fluctuations near hydrophilic surfaces are roughly Gaussian (black line), those near hydrophobic surfaces display fat low-$\tilde N$ tails.
The disparate free energetics of density fluctuations adjacent to the OH and CH$_3$ SAM surfaces are readily captured by the sparse sampling method, and agree well with those obtained from umbrella sampling; however, the sparse results were obtained at a tiny fraction of the computational cost of the umbrella sampling results.
(d) The susceptibility, $-\partial\avgNt_\phi/\partial(\beta\phi)=\varNt$, also shows distinct features for the two SAM surfaces. While the susceptibility of the OH-terminated SAM surface is roughly constant, a sharp peak is observed in the susceptibility of the CH$_3$-terminated SAM surface due to the corresponding sigmoidal response of  $\avgNt_\phi$ to $\phi$.
}
\label{fig:sam}
\end{center}
\end{figure}

\subsection{SAM -- water interfaces}
%
The low-$\tilde N$ tail of water density fluctuations, estimated in the vicinity of a surface, is expected to depend strongly on surface hydrophobicity, or the strength of surface-water interactions~\cite{LCW,mittal_pnas08,Godawat:PNAS:2009,garde09prl,Patel:JPCB:2010,LLCW,Acharya:Faraday:2010,Jamadagni:ARCB:2011,Patel:PNAS:2011,Rotenberg:JACS:2011,Patel:JPCB:2012,Patel:JPCB:2014,Remsing:JCP:2015}.
To illustrate that the sparse sampling method can capture such marked differences in the respective $F_v(\tilde N)$ profiles, here we use it to estimate $F_v(\tilde N)$ in a cylindrical disk-shaped volume, $v$, adjacent to a hydrophobic CH$_3$-terminated SAM surface (Figure~\ref{fig:sam}a) and a hydrophilic OH-terminated SAM surface;
in both cases, $v$ contains roughly $\avgNt_0\approx120$ waters on average.
As shown in Figure~\ref{fig:sam}b, the simulated $\avgNt_\phi$-values decrease with increasing $\phi$ for both the SAM surfaces; however, the decrease in $\avgNt_\phi$ is more rapid for the hydrophobic SAM surface. 
Using these $\avgNt_\phi$-values in conjunction with Equation~\ref{eq:central-phi2} allows us to estimate $F_v(\tilde N)$ adjacent to the two SAM surfaces, as shown in Figure~\ref{fig:sam}c.
While the sparse sampled estimates of $F_v(\tilde N)$ for the two SAM surfaces are very different from one another, they are nevertheless in excellent agreement with the exact results obtained from umbrella sampling.

$F_v(\tilde N)$ adjacent to the hydrophilic SAM surface is parabolic (black line) to a good approximation, consistent with the underlying density fluctuations being Gaussian.
In contrast, a marked fat tail in water density fluctuations is observed adjacent to the hydrophobic SAM surface, in agreement with previous findings~\cite{Patel:JPCB:2010,LLCW,Patel:JPCB:2012}.
Such a difference has previously been demonstrated to arise from the fact that water near hydrophobic surfaces is situated at the edge of a dewetting transition~\cite{Patel:JPCB:2012,Remsing:PNAS:2015}.
This difference is also reflected in the sensitive response of interfacial water to perturbations, that is, in the sigmoidal decrease in $\avgNt_\phi$ with increasing $\phi$ (Figure~\ref{fig:sam}b), and in a peak in the corresponding susceptibility, $-\partial\avgNt_\phi/\partial(\beta\phi)=\langle\delta \tilde{N}_v^2\rangle_\phi$ (Figure~\ref{fig:sam}d).
In contrast, the Gaussian fluctuation adjacent to the hydrophilic surface are associated with linear decrease in $\avgNt_\phi$ (Figure~\ref{fig:sam}b) and a constant susceptibility (Figure~\ref{fig:sam}d).

It is clear that there is a correspondence between the functional forms of the free energetics of water number fluctuations, $F_v(\tilde N)$, and the corresponding response of the average water number, $\avgNt_\phi$, to the strength of the potential, $\phi$.
In particular, if $F_v(\tilde N)$ is known over the entire range of $\tilde N$-values of interest, $\avgNt_\phi$ can be readily obtained at all values of $\phi$ through reweighting~\cite{Patel:JPCB:2014}.
Indeed, the curves shown in Figures~\ref{fig:sam}b and \ref{fig:sam}d were obtained in that manner.
In contrast, the sparse sampling method introduced above not only allows us to perform the inverse operation, it does so with only a select few values of $\avgNt_\phi$; that is, given well-separated $\avgNt_\phi$ estimates at a select few $\phi$-values, the sparse sampling method allows us to estimate $F_v(\tilde N)$ at those $\tilde N=\avgNt_\phi$-values.
%
%

\begin{figure}[htbp]
\begin{center}
\includegraphics[width=0.75\textwidth]{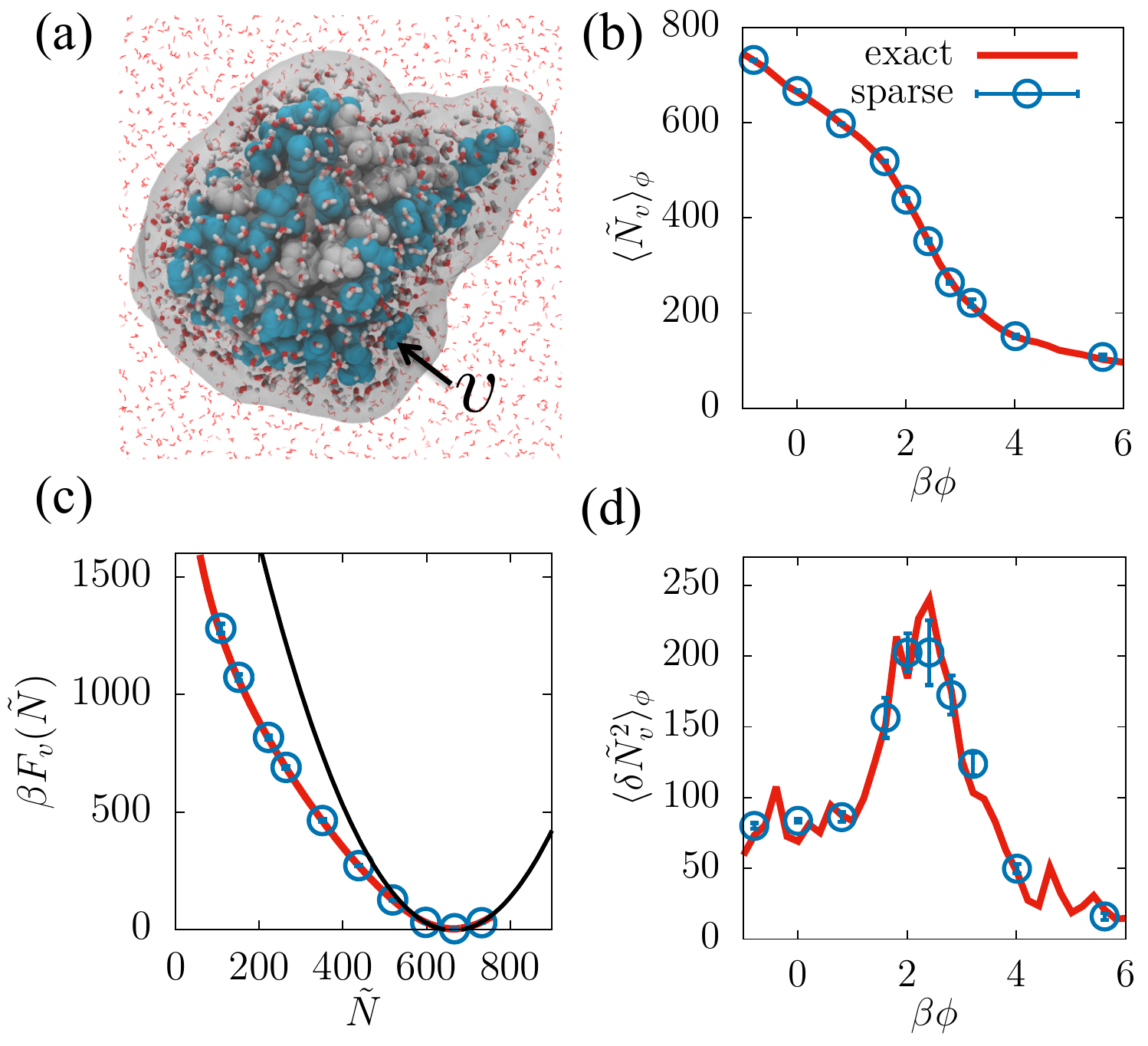}
\caption{
Estimating the free energetics of water density fluctuations in the entire hydration shell of a protein.
(a) Simulation snapshot illustrating protein in space-fill representation, with the non-polar residues colored white, and the polar and charged residues colored in blue.
Water molecules in the hydration shell (gray) are shown as sticks (red/white), while those outside the hydration shell are shown as lines.
(b) The response of the average number of coarse-grained waters, $\avgNt_\phi$, in the protein hydration shell, $v$, to the strength of the biasing potential, $\phi$.
While the response is linear near $\phi=0$, it becomes sigmoidal at higher $\phi$-values, suggesting the presence of a hydrophobic patch on the ubiquitin surface.
(c) The free energetics of water density fluctuations, $F_v(\tilde N)$, are Gaussian (black line) near the mean, but display a marked fat tail at lower $\tilde N$-values.
The sparsely sampled $F_v(\tilde N)$-values (symbols) are not only in excellent agreement with umbrella sampling (red line), but capture the functional form of $F_v(\tilde N)$ at a 42-fold reduction in the computational cost.
(d) The susceptibility, $-\partial\avgNt_\phi/\partial(\beta\phi)=\varNt$, displays a peak near $\beta \phi = 2$, suggesting a collective dewetting of water molecules from the ubiquitin hydration shell.
}
\label{fig:protein}
\end{center}
\end{figure}

\subsection{Entire Protein Hydration Shell}
%
Water density fluctuations have previously been used to characterize the hydrophobicity of regions within protein hydration shells, in volumes containing a few to roughly ten waters on average. 
Here we leverage the efficiency of the sparse sampling method to estimate water density fluctuations in the entire hydration shell of the protein, ubiquitin (Figure~\ref{fig:protein}a), which is significantly larger, and contains roughly 660 waters on average.
In contrast with the uniform SAM surfaces studied in the previous section, the protein--water interface is chemically and topographically heterogeneous; such complexity influences the corresponding water density fluctuations in a non-trivial manner~\cite{Acharya:Faraday:2010,Jamadagni:ARCB:2011}.
Using our central Equation~\ref{eq:central-phi2}, the response, $\avgNt_\phi$, to the biasing potential, $\phi \Nt$, shown in Figure~\ref{fig:protein}b, can readily be transformed into the free energy, $F_v(\tilde N)$.
As shown in Figure~\ref{fig:protein}c, $F_v(\tilde N)$ thus obtained is once again in excellent agreement with the umbrella sampling results, albeit at a fraction of the computational cost.

Interestingly, the fluctuations display a marked low-$\tilde N$ fat tail relative to Gaussian statistics (black line), suggesting the presence of extended hydrophobic regions in the hydration shell of ubiquitin. 
Indeed, ubiquitin is known to have a hydrophobic patch, which facilitates its targeting of proteins for proteasome degradation~\cite{sloper2001distinct}.
The ubiquitin hydration shell also displays a peak in the susceptibility, as shown in Figure~\ref{fig:protein}d, suggesting a collective dewetting of water molecules from the protein hydration shell.
We note certain commonalities between our results and reports of percolation transitions on partially hydrated protein surfaces.
It has been shown that when proteins are hydrated with insufficient waters to form a complete protein hydration shell, an inter-connected network of waters appears abruptly over a narrow range of hydration levels~\cite{smolin2005properties,oleinikova2005formation,Cui:Protein:2014}.
When a protein is partially hydrated, its hydration waters are essentially in vacuum; in contrast, the hydration shell waters in our simulations are surrounded by and interact with other waters.
However, in both cases, a collective (percolation or dewetting) transition is facilitated by a competition between protein-water and water-water interactions, with the protein providing a heterogeneous surface that displays a wide range of surface chemistries and thereby protein-water interaction strengths.

\subsection{Efficiency and Limitations of the Method}
%
We note that $F_v(\tilde N)$, shown in Figure~\ref{fig:protein}c, was obtained using 10 simulations run for 0.2~ns per simulation (including equilibration in response to the biasing potential) for a total simulation time of 2~ns.
In contrast, the exact umbrella sampling results employed 28 windows run for 3~ns per window for a total simulation time of 84~ns.
Thus, a dramatic speed-up in computational efficiency can be achieved if only sparse estimates of $F_v(\tilde N)$ are desired.
At the heart of this remarkable efficiency of the method is the fact that not just fewer, but shorter simulations are needed.
Because $\avgNt_\phi$ decreases monotonically with $\phi$, and can even be linear in $\phi$, $F_\phi$ can be estimated accurately with estimates of $\avgNt_\phi$ at only a few $\phi$-values.
Additionally, because the averages, $\avgNt_\phi$, are typically dominated by the most probable regions of the underlying unimodal biased distributions (Figure~\ref{fig:mult}a), they converge rapidly, and short simulations are  sufficient to accurately estimate them.
To further understand the source of the method's efficiency and for details on how to implement it optimally (for example, how to adaptively pick the set of $\phi$-values for running biased simulations), the reader is referred to ref.~\onlinecite{Patel:JPCB:2014}.
%

\begin{figure}[htbp]
\begin{center}
\includegraphics[width=0.75\textwidth]{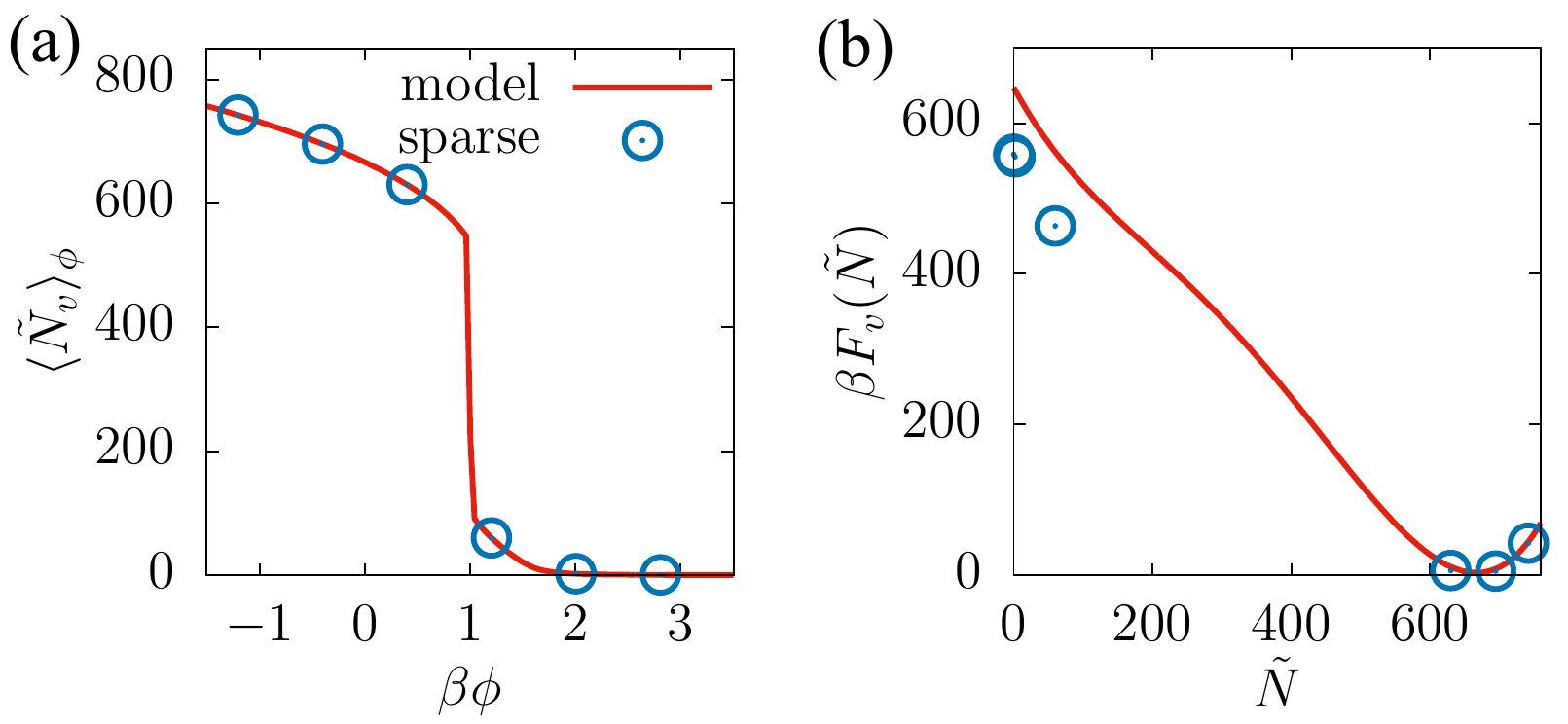}
\caption{
(a) For the analytical model discussed in the text, $\avgNt_\phi$ decreases sharply over a narrow range of $\phi$-values, making it challenging to sample intermediate $\Nt$-values. 
(b) 
The sparse sampling method displays substantive errors in $\fvnt$ for the lower $\tilde N$-values, and is unable to estimate $\fvnt$ over a wide range of intermediate $\Nt$-values.
}
\label{fig:caveat}
\end{center}
\end{figure}

The efficient estimation of $\avgNt_\phi$ relies on the most probable region(s) of the corresponding biased $\Nt$-distribution being well-sampled; this is readily achieved when the distribution is unimodal.
Because accurate estimates of $\avgNt_\phi$ are required at a number of $\phi$-values in order to obtain $F_v(\tilde N)$ accurately, it is important that every biased $\Nt$-distribution be adequately sampled.
While this is true for the results shown in Figures~\ref{fig:bulk} --~\ref{fig:protein}, it may not always be the case.
To illustrate what happens when when the biased $\Nt$-distributions are not unimodal, 
and highlight an important limitation of the sparse sampling method, we consider an analytical model for $\fvnt$, designed to yield a bimodal biased distribution for a certain biasing strength $\phi_0$, 
\begin{equation}
\fvnt = \kappa_0 (\tilde N - 100)^2 (\tilde N - 540)^2 - \phi_0 \tilde N.
\label{eq:toy-model}
\end{equation}
We choose $\beta\phi_0=1$, and to facilitate comparison with the ubiquitin $\fvnt$ (Figure~\ref{fig:protein}), we choose $\beta\kappa_0=10^{-8}$ so that the two distributions have roughly the same mean and variance. 
As shown in Figure~\ref{fig:caveat}a, $\avgNt_\phi$ for the analytical model features a sharp cliff, wherein the value of $\avgNt_\phi$ decreases dramatically over a narrow range of $\phi$-values.
The corresponding $\fvnt$ obtained using the sparse sampling method is shown in Figure~\ref{fig:caveat}b, and highlights the limitations of the method.
Due to the sharp cliff in $\avgNt_\phi$, and the associated inability to sample intermediate $\tilde N$-values, estimates of $\fvnt$ can not be obtained for a large range $\tilde N$-values.
In addition, an inability to capture the precise location of the cliff, leads to substantive errors in $F_\phi$ for higher $\phi$-values, and thereby in $\fvnt$ for the lower $\tilde N$-values.
A sharply decreasing response function akin to that for the analytical model is thus a strong indicator of bistability in the underlying free energetics; in such cases, the sparse sampling method should not be used in conjunction with linear potentials as proposed here.
Strategies for overcoming such limitations, which are likely to be encountered when water is at or close to coexistence with its vapor in the volume of interest (e.g., in hydrophobic confinement), will be the subject of a future publication.
%

\section{Outlook}
%
By circumventing the umbrella sampling requirement of overlap between adjacent biased distributions, and instead using thermodynamic integration to estimate free energy differences between the biased and unbiased ensembles, the method presented here enables sparse sampling of the free energetics of water density fluctuations, $\fvnt$.
Furthermore, a judicious choice of the functional form of the biasing potential, that is, one which is linear in the  order parameter of interest, $\Nt$, enables estimation of $\fvnt$ in a computationally efficient manner.
The low-$\tilde N$ behavior of $\fvnt$ serves to characterize the hydrophobicity of complex surfaces with nanoscale heterogeneities, such as those presented by proteins; here, we use the method to characterize $\fvnt$ in a large volume, constituting the entire hydration shell of the protein, ubiquitin.
Such a characterization of protein hydration shells not only provides an overall measure of protein hydrophobicity, but also quantifies the free energy required to displace water molecules from the protein hydration shell, which could inform its propensity to bind hydrophobic ligands~\cite{Freed:JPCB:2011}.
In addition to facilitating characterization of $\fvnt$ in large volumes, the efficiency of our method may also enable estimation of $\fvnt$, using more detailed treatments of water in the bulk or at interfaces, which are inherently expensive form a computational standpoint.
Such treatments include force fields that explicitly account for molecular polarizability~\cite{warren2008electrostatic,soniat2012effects}, or ab initio molecular dynamics simulations, which incorporate electronic structure effects~\cite{hassanali2013proton,remsing2014role}.

The method presented here is fairly general, and can be readily generalized to other order parameters besides $\Nt$ as well as to higher dimensions.
A straightforward generalization of the method could facilitate characterization of the free energetics of concentrations (as opposed to number density) fluctuations in multi-component aqueous solutions and mixtures~\cite{gupta2014structure}.
Biasing potentials that couple linearly to an order parameter of interest are also employed in a variety of other contexts; examples include constant electrostatic potential simulations used to study charge fluctuations in capacitors~\cite{Limmer:PRL:2013}, simulations that bias trajectory space using a dynamical order parameters such as activity~\cite{elmatad2010finite}, and alchemical methods for estimating binding free energies~\cite{mobley2006use,chodera2011alchemical}.
While free energy perturbation is typically employed in alchemical calculations to estimate the free energy differences between the ensembles of interest, the method introduced here could additionally provide the statistics of the energy differences between those ensembles; an understanding of such statistics may further inform optimal strategies for accurately and efficiently estimating the corresponding free energies.
Such an understanding may also lead to the development of analytical expressions for the estimation of free energies.
Due to the dramatic increase in computational efficiency that the method provides, we believe that it will also be well-suited for the characterization of free energies in multiple dimensions, wherein ensuring overlap in all order parameters becomes very expensive, and increases exponentially with the number of dimensions. 
In particular, the method may find use in the characterization of two-dimensional free energetic landscapes that serve as a starting point for recent spatial coarse-graining schemes~\cite{sinno3}.
%

\begin{acknowledgement}
A.J.P. and R.C.R. were supported in part by a Seed Grant from the University of Pennsylvania Materials Research Science and Engineering Center (NSF UPENN MRSEC DMR 11-20901).
A.J.P. also acknowledges support from the National Science Foundation (CBET 1511437), and thanks Shekhar Garde and Talid Sinno for numerous helpful discussions.
\end{acknowledgement}


\begin{mcitethebibliography}{92}
\providecommand*\natexlab[1]{#1}
\providecommand*\mciteSetBstSublistMode[1]{}
\providecommand*\mciteSetBstMaxWidthForm[2]{}
\providecommand*\mciteBstWouldAddEndPuncttrue
  {\def\EndOfBibitem{\unskip.}}
\providecommand*\mciteBstWouldAddEndPunctfalse
  {\let\EndOfBibitem\relax}
\providecommand*\mciteSetBstMidEndSepPunct[3]{}
\providecommand*\mciteSetBstSublistLabelBeginEnd[3]{}
\providecommand*\EndOfBibitem{}
\mciteSetBstSublistMode{f}
\mciteSetBstMaxWidthForm{subitem}{(\alph{mcitesubitemcount})}
\mciteSetBstSublistLabelBeginEnd
  {\mcitemaxwidthsubitemform\space}
  {\relax}
  {\relax}

\bibitem[Stillinger(1973)]{FHS:1973}
Stillinger,~F.~H. \emph{J. Solution Chem.} \textbf{1973}, \emph{2},
  141--158\relax
\mciteBstWouldAddEndPuncttrue
\mciteSetBstMidEndSepPunct{\mcitedefaultmidpunct}
{\mcitedefaultendpunct}{\mcitedefaultseppunct}\relax
\EndOfBibitem
\bibitem[Pratt and Chandler(1977)Pratt, and Chandler]{Pratt:JCP:1977}
Pratt,~L.~R.; Chandler,~D. \emph{J. Chem. Phys.} \textbf{1977}, \emph{67},
  3683--3704\relax
\mciteBstWouldAddEndPuncttrue
\mciteSetBstMidEndSepPunct{\mcitedefaultmidpunct}
{\mcitedefaultendpunct}{\mcitedefaultseppunct}\relax
\EndOfBibitem
\bibitem[Hummer \latin{et~al.}(1996)Hummer, Garde, Garcia, Pohorille, and
  Pratt]{Hummer:PNAS:1996}
Hummer,~G.; Garde,~S.; Garcia,~A.~E.; Pohorille,~A.; Pratt,~L.~R. \emph{Proc.
  Natl. Acad. Sci.} \textbf{1996}, \emph{93}, 8951--8955\relax
\mciteBstWouldAddEndPuncttrue
\mciteSetBstMidEndSepPunct{\mcitedefaultmidpunct}
{\mcitedefaultendpunct}{\mcitedefaultseppunct}\relax
\EndOfBibitem
\bibitem[Ashbaugh and Pratt(2006)Ashbaugh, and Pratt]{ashbaugh_SPT}
Ashbaugh,~H.~S.; Pratt,~L.~R. \emph{Rev. Mod. Phys.} \textbf{2006}, \emph{78},
  159\relax
\mciteBstWouldAddEndPuncttrue
\mciteSetBstMidEndSepPunct{\mcitedefaultmidpunct}
{\mcitedefaultendpunct}{\mcitedefaultseppunct}\relax
\EndOfBibitem
\bibitem[Lum \latin{et~al.}(1999)Lum, Chandler, and Weeks]{LCW}
Lum,~K.; Chandler,~D.; Weeks,~J.~D. \emph{J. Phys. Chem. B} \textbf{1999},
  \emph{103}, 4570--4577\relax
\mciteBstWouldAddEndPuncttrue
\mciteSetBstMidEndSepPunct{\mcitedefaultmidpunct}
{\mcitedefaultendpunct}{\mcitedefaultseppunct}\relax
\EndOfBibitem
\bibitem[Huang \latin{et~al.}(2001)Huang, Geissler, and Chandler]{HGC}
Huang,~D.~M.; Geissler,~P.~L.; Chandler,~D. \emph{J. Phys. Chem. B}
  \textbf{2001}, \emph{105}, 6704--6709\relax
\mciteBstWouldAddEndPuncttrue
\mciteSetBstMidEndSepPunct{\mcitedefaultmidpunct}
{\mcitedefaultendpunct}{\mcitedefaultseppunct}\relax
\EndOfBibitem
\bibitem[Huang and Chandler(2002)Huang, and Chandler]{Huang:JPCB:2002}
Huang,~D.~M.; Chandler,~D. \emph{J. Phys. Chem. B} \textbf{2002}, \emph{106},
  2047--2053\relax
\mciteBstWouldAddEndPuncttrue
\mciteSetBstMidEndSepPunct{\mcitedefaultmidpunct}
{\mcitedefaultendpunct}{\mcitedefaultseppunct}\relax
\EndOfBibitem
\bibitem[Chandler(2005)]{Chandler:Nature:2005}
Chandler,~D. \emph{Nature} \textbf{2005}, \emph{437}, 640--647\relax
\mciteBstWouldAddEndPuncttrue
\mciteSetBstMidEndSepPunct{\mcitedefaultmidpunct}
{\mcitedefaultendpunct}{\mcitedefaultseppunct}\relax
\EndOfBibitem
\bibitem[Mittal and Hummer(2008)Mittal, and Hummer]{mittal_pnas08}
Mittal,~J.; Hummer,~G. \emph{Proc. Natl. Acad. Sci.} \textbf{2008}, \emph{105},
  20130--20135\relax
\mciteBstWouldAddEndPuncttrue
\mciteSetBstMidEndSepPunct{\mcitedefaultmidpunct}
{\mcitedefaultendpunct}{\mcitedefaultseppunct}\relax
\EndOfBibitem
\bibitem[Sarupria and Garde(2009)Sarupria, and Garde]{garde09prl}
Sarupria,~S.; Garde,~S. \emph{Phys. Rev. Lett.} \textbf{2009}, \emph{103},
  037803\relax
\mciteBstWouldAddEndPuncttrue
\mciteSetBstMidEndSepPunct{\mcitedefaultmidpunct}
{\mcitedefaultendpunct}{\mcitedefaultseppunct}\relax
\EndOfBibitem
\bibitem[Godawat \latin{et~al.}(2009)Godawat, Jamadagni, and
  Garde]{Godawat:PNAS:2009}
Godawat,~R.; Jamadagni,~S.~N.; Garde,~S. \emph{Proc. Natl. Acad. Sci. U.S.A.}
  \textbf{2009}, \emph{106}, 15119 -- 15124\relax
\mciteBstWouldAddEndPuncttrue
\mciteSetBstMidEndSepPunct{\mcitedefaultmidpunct}
{\mcitedefaultendpunct}{\mcitedefaultseppunct}\relax
\EndOfBibitem
\bibitem[Acharya \latin{et~al.}(2010)Acharya, Vembanur, Jamadagni, and
  Garde]{Acharya:Faraday:2010}
Acharya,~H.; Vembanur,~S.; Jamadagni,~S.~N.; Garde,~S. \emph{Faraday Discuss.}
  \textbf{2010}, \emph{146}, 353--365\relax
\mciteBstWouldAddEndPuncttrue
\mciteSetBstMidEndSepPunct{\mcitedefaultmidpunct}
{\mcitedefaultendpunct}{\mcitedefaultseppunct}\relax
\EndOfBibitem
\bibitem[Patel \latin{et~al.}(2010)Patel, Varilly, and
  Chandler]{Patel:JPCB:2010}
Patel,~A.~J.; Varilly,~P.; Chandler,~D. \emph{J. Phys. Chem. B} \textbf{2010},
  \emph{114}, 1632 -- 1637\relax
\mciteBstWouldAddEndPuncttrue
\mciteSetBstMidEndSepPunct{\mcitedefaultmidpunct}
{\mcitedefaultendpunct}{\mcitedefaultseppunct}\relax
\EndOfBibitem
\bibitem[Patel \latin{et~al.}(2011)Patel, Varilly, Jamadagni, Acharya, Garde,
  and Chandler]{Patel:PNAS:2011}
Patel,~A.~J.; Varilly,~P.; Jamadagni,~S.~N.; Acharya,~H.; Garde,~S.;
  Chandler,~D. \emph{Proc. Natl. Acad. Sci. U.S.A.} \textbf{2011}, \emph{108},
  17678 -- 17683\relax
\mciteBstWouldAddEndPuncttrue
\mciteSetBstMidEndSepPunct{\mcitedefaultmidpunct}
{\mcitedefaultendpunct}{\mcitedefaultseppunct}\relax
\EndOfBibitem
\bibitem[Varilly \latin{et~al.}(2011)Varilly, Patel, and Chandler]{LLCW}
Varilly,~P.; Patel,~A.~J.; Chandler,~D. \emph{J. Chem. Phys.} \textbf{2011},
  \emph{134}, 074109\relax
\mciteBstWouldAddEndPuncttrue
\mciteSetBstMidEndSepPunct{\mcitedefaultmidpunct}
{\mcitedefaultendpunct}{\mcitedefaultseppunct}\relax
\EndOfBibitem
\bibitem[Jamadagni \latin{et~al.}(2011)Jamadagni, Godawat, and
  Garde]{Jamadagni:ARCB:2011}
Jamadagni,~S.~N.; Godawat,~R.; Garde,~S. \emph{Ann. Rev. Chem. Biomol. Engg.}
  \textbf{2011}, \emph{2}, 147--171\relax
\mciteBstWouldAddEndPuncttrue
\mciteSetBstMidEndSepPunct{\mcitedefaultmidpunct}
{\mcitedefaultendpunct}{\mcitedefaultseppunct}\relax
\EndOfBibitem
\bibitem[Chandler and Varilly(2012)Chandler, and Varilly]{Chandler:2012}
Chandler,~D.; Varilly,~P. Lectures on Molecular- and Nano-scale Fluctuations in
  Water. Proceeding of the International School of Physics ``Enrico Fermi''.
  2012; pp 75--111\relax
\mciteBstWouldAddEndPuncttrue
\mciteSetBstMidEndSepPunct{\mcitedefaultmidpunct}
{\mcitedefaultendpunct}{\mcitedefaultseppunct}\relax
\EndOfBibitem
\bibitem[Patel \latin{et~al.}(2012)Patel, Varilly, Jamadagni, Hagan, Chandler,
  and Garde]{Patel:JPCB:2012}
Patel,~A.~J.; Varilly,~P.; Jamadagni,~S.~N.; Hagan,~M.~F.; Chandler,~D.;
  Garde,~S. \emph{J. Phys. Chem. B} \textbf{2012}, \emph{116}, 2498 --
  2503\relax
\mciteBstWouldAddEndPuncttrue
\mciteSetBstMidEndSepPunct{\mcitedefaultmidpunct}
{\mcitedefaultendpunct}{\mcitedefaultseppunct}\relax
\EndOfBibitem
\bibitem[Remsing and Weeks(2013)Remsing, and Weeks]{Remsing:2013}
Remsing,~R.~C.; Weeks,~J.~D. \emph{J. Phys. Chem. B} \textbf{2013}, \emph{117},
  15479--15491\relax
\mciteBstWouldAddEndPuncttrue
\mciteSetBstMidEndSepPunct{\mcitedefaultmidpunct}
{\mcitedefaultendpunct}{\mcitedefaultseppunct}\relax
\EndOfBibitem
\bibitem[Patel and Garde(2014)Patel, and Garde]{Patel:JPCB:2014}
Patel,~A.~J.; Garde,~S. \emph{J. Phys. Chem. B} \textbf{2014}, \emph{118},
  1564--1573\relax
\mciteBstWouldAddEndPuncttrue
\mciteSetBstMidEndSepPunct{\mcitedefaultmidpunct}
{\mcitedefaultendpunct}{\mcitedefaultseppunct}\relax
\EndOfBibitem
\bibitem[Southall \latin{et~al.}(2002)Southall, Dill, and Haymet]{dill_rev02}
Southall,~N.~T.; Dill,~K.~A.; Haymet,~A. D.~J. \emph{J. Phys. Chem. B}
  \textbf{2002}, \emph{106}, 521--533\relax
\mciteBstWouldAddEndPuncttrue
\mciteSetBstMidEndSepPunct{\mcitedefaultmidpunct}
{\mcitedefaultendpunct}{\mcitedefaultseppunct}\relax
\EndOfBibitem
\bibitem[Dobson(2003)]{Dobson:2003}
Dobson,~C.~M. \emph{Nature} \textbf{2003}, \emph{426}, 884--890\relax
\mciteBstWouldAddEndPuncttrue
\mciteSetBstMidEndSepPunct{\mcitedefaultmidpunct}
{\mcitedefaultendpunct}{\mcitedefaultseppunct}\relax
\EndOfBibitem
\bibitem[Levy and Onuchic(2006)Levy, and Onuchic]{Levy:2006aa}
Levy,~Y.; Onuchic,~J.~N. \emph{Annu Rev Biophys Biomol Struct} \textbf{2006},
  \emph{35}, 389--415\relax
\mciteBstWouldAddEndPuncttrue
\mciteSetBstMidEndSepPunct{\mcitedefaultmidpunct}
{\mcitedefaultendpunct}{\mcitedefaultseppunct}\relax
\EndOfBibitem
\bibitem[Krone \latin{et~al.}(2008)Krone, Hua, Soto, Zhou, Berne, and
  Shea]{shea08}
Krone,~M.~G.; Hua,~L.; Soto,~P.; Zhou,~R.; Berne,~B.~J.; Shea,~J.-E. \emph{J.
  Am. Chem. Soc.} \textbf{2008}, \emph{130}, 11066--11072\relax
\mciteBstWouldAddEndPuncttrue
\mciteSetBstMidEndSepPunct{\mcitedefaultmidpunct}
{\mcitedefaultendpunct}{\mcitedefaultseppunct}\relax
\EndOfBibitem
\bibitem[Ball(2008)]{ball08}
Ball,~P. \emph{Chem. Rev.} \textbf{2008}, \emph{108}, 74--108\relax
\mciteBstWouldAddEndPuncttrue
\mciteSetBstMidEndSepPunct{\mcitedefaultmidpunct}
{\mcitedefaultendpunct}{\mcitedefaultseppunct}\relax
\EndOfBibitem
\bibitem[Thirumalai \latin{et~al.}(2012)Thirumalai, Reddy, and
  Straub]{Thirumalai:2012}
Thirumalai,~D.; Reddy,~G.; Straub,~J.~E. \emph{Acc. Chem Res.} \textbf{2012},
  \emph{45}, 83--92\relax
\mciteBstWouldAddEndPuncttrue
\mciteSetBstMidEndSepPunct{\mcitedefaultmidpunct}
{\mcitedefaultendpunct}{\mcitedefaultseppunct}\relax
\EndOfBibitem
\bibitem[Tanford(1973)]{Tanford1973}
Tanford,~C. \emph{The Hydrophobic Effect: Formation of Micelles and Biological
  Membranes}; John Wiley \&\ Sons, 1973\relax
\mciteBstWouldAddEndPuncttrue
\mciteSetBstMidEndSepPunct{\mcitedefaultmidpunct}
{\mcitedefaultendpunct}{\mcitedefaultseppunct}\relax
\EndOfBibitem
\bibitem[Whitesides and Grzybowski(2002)Whitesides, and
  Grzybowski]{Whitesides:2002aa}
Whitesides,~G.~M.; Grzybowski,~B. \emph{Science} \textbf{2002}, \emph{295},
  2418--21\relax
\mciteBstWouldAddEndPuncttrue
\mciteSetBstMidEndSepPunct{\mcitedefaultmidpunct}
{\mcitedefaultendpunct}{\mcitedefaultseppunct}\relax
\EndOfBibitem
\bibitem[Rabani \latin{et~al.}(2003)Rabani, Reichman, Geissler, and
  Brus]{Rabani:2003}
Rabani,~E.; Reichman,~D.~R.; Geissler,~P.~L.; Brus,~L.~E. \emph{Nature}
  \textbf{2003}, \emph{426}, 271--274\relax
\mciteBstWouldAddEndPuncttrue
\mciteSetBstMidEndSepPunct{\mcitedefaultmidpunct}
{\mcitedefaultendpunct}{\mcitedefaultseppunct}\relax
\EndOfBibitem
\bibitem[Maibaum \latin{et~al.}(2004)Maibaum, Dinner, and
  Chandler]{MaibaumDinnerChandler2004}
Maibaum,~L.; Dinner,~A.~R.; Chandler,~D. \emph{J. Phys. Chem. B} \textbf{2004},
  \emph{108}, 6778--6781\relax
\mciteBstWouldAddEndPuncttrue
\mciteSetBstMidEndSepPunct{\mcitedefaultmidpunct}
{\mcitedefaultendpunct}{\mcitedefaultseppunct}\relax
\EndOfBibitem
\bibitem[Morrone \latin{et~al.}(2012)Morrone, Li, and Berne]{Morrone:2012}
Morrone,~J.~A.; Li,~J.; Berne,~B.~J. \emph{J. Phys. Chem. B} \textbf{2012},
  \emph{116}, 378--389\relax
\mciteBstWouldAddEndPuncttrue
\mciteSetBstMidEndSepPunct{\mcitedefaultmidpunct}
{\mcitedefaultendpunct}{\mcitedefaultseppunct}\relax
\EndOfBibitem
\bibitem[Garde \latin{et~al.}(1996)Garde, Hummer, Garcia, Paulaitis, and
  Pratt]{Garde:PRL:1996}
Garde,~S.; Hummer,~G.; Garcia,~A.~E.; Paulaitis,~M.~E.; Pratt,~L.~R.
  \emph{Phys. Rev. Lett.} \textbf{1996}, \emph{77}, 4966--4968\relax
\mciteBstWouldAddEndPuncttrue
\mciteSetBstMidEndSepPunct{\mcitedefaultmidpunct}
{\mcitedefaultendpunct}{\mcitedefaultseppunct}\relax
\EndOfBibitem
\bibitem[ten Wolde(2002)]{tW_rev}
ten Wolde,~P.~R. \emph{J. Phys. Cond. Matt.} \textbf{2002}, \emph{14},
  9445--9460\relax
\mciteBstWouldAddEndPuncttrue
\mciteSetBstMidEndSepPunct{\mcitedefaultmidpunct}
{\mcitedefaultendpunct}{\mcitedefaultseppunct}\relax
\EndOfBibitem
\bibitem[Pratt(2002)]{pratt_rev}
Pratt,~L.~R. \emph{Ann. Rev. Phys. Chem.} \textbf{2002}, \emph{53},
  409--436\relax
\mciteBstWouldAddEndPuncttrue
\mciteSetBstMidEndSepPunct{\mcitedefaultmidpunct}
{\mcitedefaultendpunct}{\mcitedefaultseppunct}\relax
\EndOfBibitem
\bibitem[Hummer \latin{et~al.}(1998)Hummer, Garde, Garcia, Paulaitis, and
  Pratt]{Hummer:PNAS:1998}
Hummer,~G.; Garde,~S.; Garcia,~A.; Paulaitis,~M.; Pratt,~L. \emph{Proc. Natl.
  Acad. Sci. USA} \textbf{1998}, \emph{95}, 1552--1555\relax
\mciteBstWouldAddEndPuncttrue
\mciteSetBstMidEndSepPunct{\mcitedefaultmidpunct}
{\mcitedefaultendpunct}{\mcitedefaultseppunct}\relax
\EndOfBibitem
\bibitem[Huang and Chandler(2000)Huang, and Chandler]{HuangChandlerPRE}
Huang,~D.~M.; Chandler,~D. \emph{Phys. Rev. E} \textbf{2000}, \emph{61},
  1501--1506\relax
\mciteBstWouldAddEndPuncttrue
\mciteSetBstMidEndSepPunct{\mcitedefaultmidpunct}
{\mcitedefaultendpunct}{\mcitedefaultseppunct}\relax
\EndOfBibitem
\bibitem[Remsing and Patel(2015)Remsing, and Patel]{Remsing:JCP:2015}
Remsing,~R.~C.; Patel,~A.~J. \emph{J. Chem. Phys.} \textbf{2015}, \emph{142},
  024502\relax
\mciteBstWouldAddEndPuncttrue
\mciteSetBstMidEndSepPunct{\mcitedefaultmidpunct}
{\mcitedefaultendpunct}{\mcitedefaultseppunct}\relax
\EndOfBibitem
\bibitem[Kitchen \latin{et~al.}(1992)Kitchen, Reed, and Levy]{Levy:1992}
Kitchen,~D.~B.; Reed,~L.~H.; Levy,~R.~M. \emph{Biochemistry} \textbf{1992},
  \emph{31}, 10083--10093\relax
\mciteBstWouldAddEndPuncttrue
\mciteSetBstMidEndSepPunct{\mcitedefaultmidpunct}
{\mcitedefaultendpunct}{\mcitedefaultseppunct}\relax
\EndOfBibitem
\bibitem[Zhou \latin{et~al.}(2004)Zhou, Huang, Margulis, and Berne]{berne04}
Zhou,~R.; Huang,~X.; Margulis,~C.~J.; Berne,~B.~J. \emph{Science}
  \textbf{2004}, \emph{305}, 1605--1609\relax
\mciteBstWouldAddEndPuncttrue
\mciteSetBstMidEndSepPunct{\mcitedefaultmidpunct}
{\mcitedefaultendpunct}{\mcitedefaultseppunct}\relax
\EndOfBibitem
\bibitem[Liu \latin{et~al.}(2005)Liu, Huang, Zhou, and Berne]{berne05_melittin}
Liu,~P.; Huang,~X.; Zhou,~R.; Berne,~B.~J. \emph{Nature} \textbf{2005},
  \emph{437}, 159--162\relax
\mciteBstWouldAddEndPuncttrue
\mciteSetBstMidEndSepPunct{\mcitedefaultmidpunct}
{\mcitedefaultendpunct}{\mcitedefaultseppunct}\relax
\EndOfBibitem
\bibitem[Choudhury and Pettitt(2007)Choudhury, and Pettitt]{chou_dewet}
Choudhury,~N.; Pettitt,~B.~M. \emph{J. Am. Chem. Soc.} \textbf{2007},
  \emph{129}, 4847--4852\relax
\mciteBstWouldAddEndPuncttrue
\mciteSetBstMidEndSepPunct{\mcitedefaultmidpunct}
{\mcitedefaultendpunct}{\mcitedefaultseppunct}\relax
\EndOfBibitem
\bibitem[Choudhury and Pettitt(2005)Choudhury, and Pettitt]{chou05}
Choudhury,~N.; Pettitt,~B.~M. \emph{Mol. Sim.} \textbf{2005}, \emph{31},
  457--463\relax
\mciteBstWouldAddEndPuncttrue
\mciteSetBstMidEndSepPunct{\mcitedefaultmidpunct}
{\mcitedefaultendpunct}{\mcitedefaultseppunct}\relax
\EndOfBibitem
\bibitem[Rasaiah \latin{et~al.}(2008)Rasaiah, Garde, and Hummer]{garde_rev}
Rasaiah,~J.~C.; Garde,~S.; Hummer,~G. \emph{Ann. Rev. Phys. Chem.}
  \textbf{2008}, \emph{59}, 713--740\relax
\mciteBstWouldAddEndPuncttrue
\mciteSetBstMidEndSepPunct{\mcitedefaultmidpunct}
{\mcitedefaultendpunct}{\mcitedefaultseppunct}\relax
\EndOfBibitem
\bibitem[Vembanur \latin{et~al.}(2013)Vembanur, Patel, Sarupria, and
  Garde]{Vembanur:2013}
Vembanur,~S.; Patel,~A.~J.; Sarupria,~S.; Garde,~S. \emph{J. Phys. Chem. B}
  \textbf{2013}, \emph{117}, 10261--10270\relax
\mciteBstWouldAddEndPuncttrue
\mciteSetBstMidEndSepPunct{\mcitedefaultmidpunct}
{\mcitedefaultendpunct}{\mcitedefaultseppunct}\relax
\EndOfBibitem
\bibitem[Granick and Bae(2008)Granick, and Bae]{Granick:Science:2008}
Granick,~S.; Bae,~S.~C. \emph{Science} \textbf{2008}, \emph{322},
  1477--1478\relax
\mciteBstWouldAddEndPuncttrue
\mciteSetBstMidEndSepPunct{\mcitedefaultmidpunct}
{\mcitedefaultendpunct}{\mcitedefaultseppunct}\relax
\EndOfBibitem
\bibitem[Siebert and Hummer(2002)Siebert, and Hummer]{Siebert:Biochem:2002}
Siebert,~X.; Hummer,~G. \emph{Biochemistry} \textbf{2002}, \emph{41},
  2956--2961\relax
\mciteBstWouldAddEndPuncttrue
\mciteSetBstMidEndSepPunct{\mcitedefaultmidpunct}
{\mcitedefaultendpunct}{\mcitedefaultseppunct}\relax
\EndOfBibitem
\bibitem[Giovambattista \latin{et~al.}(2008)Giovambattista, Lopez, Rossky, and
  Debenedetti]{Giovambattista:PNAS:2008}
Giovambattista,~N.; Lopez,~C.~F.; Rossky,~P.~J.; Debenedetti,~P.~G. \emph{Proc.
  Natl. Acad. Sci. U.S.A.} \textbf{2008}, \emph{105}, 2274--2279\relax
\mciteBstWouldAddEndPuncttrue
\mciteSetBstMidEndSepPunct{\mcitedefaultmidpunct}
{\mcitedefaultendpunct}{\mcitedefaultseppunct}\relax
\EndOfBibitem
\bibitem[Mittal and Hummer(2010)Mittal, and Hummer]{Mittal:Faraday:2010}
Mittal,~J.; Hummer,~G. \emph{Faraday Discuss.} \textbf{2010}, \emph{146},
  341--352\relax
\mciteBstWouldAddEndPuncttrue
\mciteSetBstMidEndSepPunct{\mcitedefaultmidpunct}
{\mcitedefaultendpunct}{\mcitedefaultseppunct}\relax
\EndOfBibitem
\bibitem[Daub \latin{et~al.}(2010)Daub, Wang, Kudesia, Bratko, and
  Luzar]{Luzar:Faraday:2010}
Daub,~C.~D.; Wang,~J.; Kudesia,~S.; Bratko,~D.; Luzar,~A. \emph{Faraday
  Discuss.} \textbf{2010}, \emph{146}, 67--77\relax
\mciteBstWouldAddEndPuncttrue
\mciteSetBstMidEndSepPunct{\mcitedefaultmidpunct}
{\mcitedefaultendpunct}{\mcitedefaultseppunct}\relax
\EndOfBibitem
\bibitem[Giovambattista \latin{et~al.}(2007)Giovambattista, Debenedetti, and
  Rossky]{Giovambattista:JPCC:2007}
Giovambattista,~N.; Debenedetti,~P.~G.; Rossky,~P.~J. \emph{J. Phys. Chem. C}
  \textbf{2007}, \emph{111}, 1323--1332\relax
\mciteBstWouldAddEndPuncttrue
\mciteSetBstMidEndSepPunct{\mcitedefaultmidpunct}
{\mcitedefaultendpunct}{\mcitedefaultseppunct}\relax
\EndOfBibitem
\bibitem[Hua \latin{et~al.}(2009)Hua, Zangi, and Berne]{Berne:JPCC:2009}
Hua,~L.; Zangi,~R.; Berne,~B.~J. \emph{J. Phys. Chem. C} \textbf{2009},
  \emph{113}, 5244--5253\relax
\mciteBstWouldAddEndPuncttrue
\mciteSetBstMidEndSepPunct{\mcitedefaultmidpunct}
{\mcitedefaultendpunct}{\mcitedefaultseppunct}\relax
\EndOfBibitem
\bibitem[Argyris \latin{et~al.}(2009)Argyris, Cole, and
  Striolo]{Striolo:Langmuir:2009}
Argyris,~D.; Cole,~D.~R.; Striolo,~A. \emph{Langmuir} \textbf{2009}, \emph{25},
  8025--8035\relax
\mciteBstWouldAddEndPuncttrue
\mciteSetBstMidEndSepPunct{\mcitedefaultmidpunct}
{\mcitedefaultendpunct}{\mcitedefaultseppunct}\relax
\EndOfBibitem
\bibitem[Wang \latin{et~al.}(2011)Wang, Bratko, and Luzar]{Luzar:PNAS:2011}
Wang,~J.; Bratko,~D.; Luzar,~A. \emph{Proc. Natl. Acad. Sci. U.S.A.}
  \textbf{2011}, \emph{108}, 6374--6379\relax
\mciteBstWouldAddEndPuncttrue
\mciteSetBstMidEndSepPunct{\mcitedefaultmidpunct}
{\mcitedefaultendpunct}{\mcitedefaultseppunct}\relax
\EndOfBibitem
\bibitem[Giovambattista \latin{et~al.}(2009)Giovambattista, Debenedetti, and
  Rossky]{Giovambattista:PNAS:2009}
Giovambattista,~N.; Debenedetti,~P.~G.; Rossky,~P.~J. \emph{Proc. Natl. Acad.
  Sci. U.S.A.} \textbf{2009}, \emph{106}, 15181--15185\relax
\mciteBstWouldAddEndPuncttrue
\mciteSetBstMidEndSepPunct{\mcitedefaultmidpunct}
{\mcitedefaultendpunct}{\mcitedefaultseppunct}\relax
\EndOfBibitem
\bibitem[Surblys \latin{et~al.}(2011)Surblys, Yamaguchi, Kuroda, Nakajima, and
  Fujimura]{Surblys:JCP:2011}
Surblys,~D.; Yamaguchi,~Y.; Kuroda,~K.; Nakajima,~T.; Fujimura,~H. \emph{J.
  Chem. Phys.} \textbf{2011}, \emph{135}, 014703\relax
\mciteBstWouldAddEndPuncttrue
\mciteSetBstMidEndSepPunct{\mcitedefaultmidpunct}
{\mcitedefaultendpunct}{\mcitedefaultseppunct}\relax
\EndOfBibitem
\bibitem[Rotenberg \latin{et~al.}(2011)Rotenberg, Patel, and
  Chandler]{Rotenberg:JACS:2011}
Rotenberg,~B.; Patel,~A.~J.; Chandler,~D. \emph{J. Am. Chem. Soc.}
  \textbf{2011}, \emph{133}, 20521 -- 20527\relax
\mciteBstWouldAddEndPuncttrue
\mciteSetBstMidEndSepPunct{\mcitedefaultmidpunct}
{\mcitedefaultendpunct}{\mcitedefaultseppunct}\relax
\EndOfBibitem
\bibitem[Widom(1963)]{Widom:JCP:1963}
Widom,~B. \emph{J. Chem. Phys.} \textbf{1963}, \emph{39}, 2808 -- 2812\relax
\mciteBstWouldAddEndPuncttrue
\mciteSetBstMidEndSepPunct{\mcitedefaultmidpunct}
{\mcitedefaultendpunct}{\mcitedefaultseppunct}\relax
\EndOfBibitem
\bibitem[Chandler(1987)]{DCbook}
Chandler,~D. \emph{Introduction to Modern Statistical Mechanics}; Oxford
  University Press, 1987\relax
\mciteBstWouldAddEndPuncttrue
\mciteSetBstMidEndSepPunct{\mcitedefaultmidpunct}
{\mcitedefaultendpunct}{\mcitedefaultseppunct}\relax
\EndOfBibitem
\bibitem[Patel \latin{et~al.}(2011)Patel, Varilly, Chandler, and
  Garde]{Patel:JSP:2011}
Patel,~A.~J.; Varilly,~P.; Chandler,~D.; Garde,~S. \emph{J. Stat. Phys.}
  \textbf{2011}, \emph{145}, 265 -- 275\relax
\mciteBstWouldAddEndPuncttrue
\mciteSetBstMidEndSepPunct{\mcitedefaultmidpunct}
{\mcitedefaultendpunct}{\mcitedefaultseppunct}\relax
\EndOfBibitem
\bibitem[Warren and Patel(2008)Warren, and Patel]{warren2008electrostatic}
Warren,~G.~L.; Patel,~S. \emph{J. Phys. Chem. B} \textbf{2008}, \emph{112},
  11679--11693\relax
\mciteBstWouldAddEndPuncttrue
\mciteSetBstMidEndSepPunct{\mcitedefaultmidpunct}
{\mcitedefaultendpunct}{\mcitedefaultseppunct}\relax
\EndOfBibitem
\bibitem[Remsing \latin{et~al.}(2014)Remsing, Baer, Schenter, Mundy, and
  Weeks]{remsing2014role}
Remsing,~R.~C.; Baer,~M.~D.; Schenter,~G.~K.; Mundy,~C.~J.; Weeks,~J.~D.
  \emph{The Journal of Physical Chemistry Letters} \textbf{2014}, \emph{5},
  2767--2774\relax
\mciteBstWouldAddEndPuncttrue
\mciteSetBstMidEndSepPunct{\mcitedefaultmidpunct}
{\mcitedefaultendpunct}{\mcitedefaultseppunct}\relax
\EndOfBibitem
\bibitem[Frenkel and Smit(2002)Frenkel, and Smit]{Frenkel_Smit}
Frenkel,~D.; Smit,~B. \emph{Understanding Molecular Simulations: From
  Algorithms to Applications}, 2nd ed.; Academic Press, New York, 2002\relax
\mciteBstWouldAddEndPuncttrue
\mciteSetBstMidEndSepPunct{\mcitedefaultmidpunct}
{\mcitedefaultendpunct}{\mcitedefaultseppunct}\relax
\EndOfBibitem
\bibitem[Souaille and Roux(2001)Souaille, and Roux]{roux_wham}
Souaille,~M.; Roux,~B. \emph{Computer Phys. Comm.} \textbf{2001}, \emph{135},
  40 -- 57\relax
\mciteBstWouldAddEndPuncttrue
\mciteSetBstMidEndSepPunct{\mcitedefaultmidpunct}
{\mcitedefaultendpunct}{\mcitedefaultseppunct}\relax
\EndOfBibitem
\bibitem[Kumar \latin{et~al.}(1992)Kumar, Rosenberg, Bouzida, Swendsen, and
  Kollman]{wham2}
Kumar,~S.; Rosenberg,~J.~M.; Bouzida,~D.; Swendsen,~R.~H.; Kollman,~P.~A.
  \emph{J. Comp. Chem.} \textbf{1992}, \emph{13}, 1011--1021\relax
\mciteBstWouldAddEndPuncttrue
\mciteSetBstMidEndSepPunct{\mcitedefaultmidpunct}
{\mcitedefaultendpunct}{\mcitedefaultseppunct}\relax
\EndOfBibitem
\bibitem[Shirts and Chodera(2008)Shirts, and Chodera]{MBAR}
Shirts,~M.~R.; Chodera,~J.~D. \emph{J. Chem. Phys.} \textbf{2008}, \emph{129},
  124105\relax
\mciteBstWouldAddEndPuncttrue
\mciteSetBstMidEndSepPunct{\mcitedefaultmidpunct}
{\mcitedefaultendpunct}{\mcitedefaultseppunct}\relax
\EndOfBibitem
\bibitem[K{\"a}stner and Thiel(2005)K{\"a}stner, and Thiel]{Kastner:JCP:2005}
K{\"a}stner,~J.; Thiel,~W. \emph{J. Chem. Phys.} \textbf{2005}, \emph{123},
  144104\relax
\mciteBstWouldAddEndPuncttrue
\mciteSetBstMidEndSepPunct{\mcitedefaultmidpunct}
{\mcitedefaultendpunct}{\mcitedefaultseppunct}\relax
\EndOfBibitem
\bibitem[Vijay-Kumar \latin{et~al.}(1987)Vijay-Kumar, Bugg, and
  Cook]{ubiquitin}
Vijay-Kumar,~S.; Bugg,~C.~E.; Cook,~W.~J. \emph{Journal of Molecular Biology}
  \textbf{1987}, \emph{194}, 531 -- 544\relax
\mciteBstWouldAddEndPuncttrue
\mciteSetBstMidEndSepPunct{\mcitedefaultmidpunct}
{\mcitedefaultendpunct}{\mcitedefaultseppunct}\relax
\EndOfBibitem
\bibitem[Hess \latin{et~al.}(2008)Hess, Kutzner, van~der Spoel, and
  Lindahl]{gmx4ref}
Hess,~B.; Kutzner,~C.; van~der Spoel,~D.; Lindahl,~E. \emph{J. Chem. Theory
  Comp.} \textbf{2008}, 435 -- 447\relax
\mciteBstWouldAddEndPuncttrue
\mciteSetBstMidEndSepPunct{\mcitedefaultmidpunct}
{\mcitedefaultendpunct}{\mcitedefaultseppunct}\relax
\EndOfBibitem
\bibitem[Berendsen \latin{et~al.}(1987)Berendsen, Grigera, and Straatsma]{SPCE}
Berendsen,~H. J.~C.; Grigera,~J.~R.; Straatsma,~T.~P. \emph{J. Phys. Chem.}
  \textbf{1987}, \emph{91}, 6269--6271\relax
\mciteBstWouldAddEndPuncttrue
\mciteSetBstMidEndSepPunct{\mcitedefaultmidpunct}
{\mcitedefaultendpunct}{\mcitedefaultseppunct}\relax
\EndOfBibitem
\bibitem[Jorgensen \latin{et~al.}(1983)Jorgensen, Chandrasekhar, Madura, Impey,
  and Klein]{TIP4P}
Jorgensen,~W.~L.; Chandrasekhar,~J.; Madura,~J.~D.; Impey,~R.~W.; Klein,~M.~L.
  \emph{J. Chem. Phys.} \textbf{1983}, \emph{79}, 926--935\relax
\mciteBstWouldAddEndPuncttrue
\mciteSetBstMidEndSepPunct{\mcitedefaultmidpunct}
{\mcitedefaultendpunct}{\mcitedefaultseppunct}\relax
\EndOfBibitem
\bibitem[Essmann \latin{et~al.}(1995)Essmann, Perera, Berkowitz, Darden, Lee,
  and Pedersen]{PME}
Essmann,~U.; Perera,~L.; Berkowitz,~M.~L.; Darden,~T.; Lee,~H.; Pedersen,~L.~G.
  \emph{J. Chem. Phys.} \textbf{1995}, \emph{103}, 8577--8593\relax
\mciteBstWouldAddEndPuncttrue
\mciteSetBstMidEndSepPunct{\mcitedefaultmidpunct}
{\mcitedefaultendpunct}{\mcitedefaultseppunct}\relax
\EndOfBibitem
\bibitem[Ryckaert \latin{et~al.}(1977)Ryckaert, Ciccotti, and Berendsen]{SHAKE}
Ryckaert,~J.-P.; Ciccotti,~G.; Berendsen,~H. J.~C. \emph{J. Comp. Phys.}
  \textbf{1977}, \emph{23}, 327 -- 341\relax
\mciteBstWouldAddEndPuncttrue
\mciteSetBstMidEndSepPunct{\mcitedefaultmidpunct}
{\mcitedefaultendpunct}{\mcitedefaultseppunct}\relax
\EndOfBibitem
\bibitem[Hess(2008)]{LINCS}
Hess,~B. \emph{J. Chem. Theory Comput.} \textbf{2008}, \emph{4}, 116--122\relax
\mciteBstWouldAddEndPuncttrue
\mciteSetBstMidEndSepPunct{\mcitedefaultmidpunct}
{\mcitedefaultendpunct}{\mcitedefaultseppunct}\relax
\EndOfBibitem
\bibitem[Bussi \latin{et~al.}(2007)Bussi, Donadio, and Parrinello]{V-Rescale}
Bussi,~G.; Donadio,~D.; Parrinello,~M. \emph{J. Chem. Phys.} \textbf{2007},
  \emph{126}, 014101\relax
\mciteBstWouldAddEndPuncttrue
\mciteSetBstMidEndSepPunct{\mcitedefaultmidpunct}
{\mcitedefaultendpunct}{\mcitedefaultseppunct}\relax
\EndOfBibitem
\bibitem[Parrinello and Rahman()Parrinello, and Rahman]{Parrinello-Rahman}
Parrinello,~M.; Rahman,~A. \emph{J. Applied Phys.} \emph{52}, 7182--7190\relax
\mciteBstWouldAddEndPuncttrue
\mciteSetBstMidEndSepPunct{\mcitedefaultmidpunct}
{\mcitedefaultendpunct}{\mcitedefaultseppunct}\relax
\EndOfBibitem
\bibitem[Miller \latin{et~al.}(2007)Miller, Vanden-Eijnden, and
  Chandler]{Miller:PNAS:2007}
Miller,~T.; Vanden-Eijnden,~E.; Chandler,~D. \emph{Proc. Natl. Acad. Sci.
  U.S.A.} \textbf{2007}, \emph{104}, 14559--14564\relax
\mciteBstWouldAddEndPuncttrue
\mciteSetBstMidEndSepPunct{\mcitedefaultmidpunct}
{\mcitedefaultendpunct}{\mcitedefaultseppunct}\relax
\EndOfBibitem
\bibitem[MacKerell \latin{et~al.}(1998)MacKerell, Bashford, Bellott, Dunbrack,
  Evanseck, Field, Fischer, Gao, Guo, Ha, Joseph-McCarthy, Kuchnir, Kuczera,
  Lau, Mattos, Michnick, Ngo, Nguyen, Prodhom, Reiher, Roux, Schlenkrich,
  Smith, Stote, Straub, Watanabe, Wi{\'o}rkiewicz-Kuczera, Yin, and
  Karplus]{CHARMM}
MacKerell,~A.~D.; Bashford,~D.; Bellott,~M.; Dunbrack,~R.~L.; Evanseck,~J.~D.;
  Field,~M.~J.; Fischer,~S.; Gao,~J.; Guo,~H.; Ha,~S.; Joseph-McCarthy,~D.;
  Kuchnir,~L.; Kuczera,~K.; Lau,~F. T.~K.; Mattos,~C.; Michnick,~S.; Ngo,~T.;
  Nguyen,~D.~T.; Prodhom,~B.; Reiher,~W.~E.; Roux,~B.; Schlenkrich,~M.;
  Smith,~J.~C.; Stote,~R.; Straub,~J.; Watanabe,~M.;
  Wi{\'o}rkiewicz-Kuczera,~J.; Yin,~D.; Karplus,~M. \emph{J. Phys. Chem. B}
  \textbf{1998}, \emph{102}, 3586--3616\relax
\mciteBstWouldAddEndPuncttrue
\mciteSetBstMidEndSepPunct{\mcitedefaultmidpunct}
{\mcitedefaultendpunct}{\mcitedefaultseppunct}\relax
\EndOfBibitem
\bibitem[Remsing \latin{et~al.}(2015)Remsing, Xi, Vembanur, Sharma,
  Debenedetti, Garde, and Patel]{Remsing:PNAS:2015}
Remsing,~R.~C.; Xi,~E.; Vembanur,~S.; Sharma,~S.; Debenedetti,~P.~G.;
  Garde,~S.; Patel,~A.~J. \emph{Proc. Natl. Acad. Sci. U.S.A.} \textbf{2015},
  \emph{112}, 8181--8186\relax
\mciteBstWouldAddEndPuncttrue
\mciteSetBstMidEndSepPunct{\mcitedefaultmidpunct}
{\mcitedefaultendpunct}{\mcitedefaultseppunct}\relax
\EndOfBibitem
\bibitem[Sloper-Mould \latin{et~al.}(2001)Sloper-Mould, Jemc, Pickart, and
  Hicke]{sloper2001distinct}
Sloper-Mould,~K.~E.; Jemc,~J.~C.; Pickart,~C.~M.; Hicke,~L. \emph{J. Biol.
  Chem.} \textbf{2001}, \emph{276}, 30483--30489\relax
\mciteBstWouldAddEndPuncttrue
\mciteSetBstMidEndSepPunct{\mcitedefaultmidpunct}
{\mcitedefaultendpunct}{\mcitedefaultseppunct}\relax
\EndOfBibitem
\bibitem[Smolin \latin{et~al.}(2005)Smolin, Oleinikova, Brovchenko, Geiger, and
  Winter]{smolin2005properties}
Smolin,~N.; Oleinikova,~A.; Brovchenko,~I.; Geiger,~A.; Winter,~R. \emph{J.
  Phys. Chem. B} \textbf{2005}, \emph{109}, 10995--11005\relax
\mciteBstWouldAddEndPuncttrue
\mciteSetBstMidEndSepPunct{\mcitedefaultmidpunct}
{\mcitedefaultendpunct}{\mcitedefaultseppunct}\relax
\EndOfBibitem
\bibitem[Oleinikova \latin{et~al.}(2005)Oleinikova, Smolin, Brovchenko, Geiger,
  and Winter]{oleinikova2005formation}
Oleinikova,~A.; Smolin,~N.; Brovchenko,~I.; Geiger,~A.; Winter,~R. \emph{J.
  Phys. Chem. B} \textbf{2005}, \emph{109}, 1988--1998\relax
\mciteBstWouldAddEndPuncttrue
\mciteSetBstMidEndSepPunct{\mcitedefaultmidpunct}
{\mcitedefaultendpunct}{\mcitedefaultseppunct}\relax
\EndOfBibitem
\bibitem[Cui \latin{et~al.}(2014)Cui, Ou, and Patel]{Cui:Protein:2014}
Cui,~D.; Ou,~S.; Patel,~S. \emph{Proteins: Struc. Func. Bioinformatics}
  \textbf{2014}, \emph{82}, 3312--3326\relax
\mciteBstWouldAddEndPuncttrue
\mciteSetBstMidEndSepPunct{\mcitedefaultmidpunct}
{\mcitedefaultendpunct}{\mcitedefaultseppunct}\relax
\EndOfBibitem
\bibitem[Freed \latin{et~al.}(2011)Freed, Garde, and Cramer]{Freed:JPCB:2011}
Freed,~A.~S.; Garde,~S.; Cramer,~S.~M. \emph{J. Phys. Chem. B} \textbf{2011},
  \emph{115}, 13320--13327\relax
\mciteBstWouldAddEndPuncttrue
\mciteSetBstMidEndSepPunct{\mcitedefaultmidpunct}
{\mcitedefaultendpunct}{\mcitedefaultseppunct}\relax
\EndOfBibitem
\bibitem[Soniat and Rick(2012)Soniat, and Rick]{soniat2012effects}
Soniat,~M.; Rick,~S.~W. \emph{J. Chem. Phys.} \textbf{2012}, \emph{137},
  044511\relax
\mciteBstWouldAddEndPuncttrue
\mciteSetBstMidEndSepPunct{\mcitedefaultmidpunct}
{\mcitedefaultendpunct}{\mcitedefaultseppunct}\relax
\EndOfBibitem
\bibitem[Hassanali \latin{et~al.}(2013)Hassanali, Giberti, Cuny, K{\"u}hne, and
  Parrinello]{hassanali2013proton}
Hassanali,~A.; Giberti,~F.; Cuny,~J.; K{\"u}hne,~T.~D.; Parrinello,~M.
  \emph{Proc. Natl. Acad. Sci. U.S.A.} \textbf{2013}, \emph{110},
  13723--13728\relax
\mciteBstWouldAddEndPuncttrue
\mciteSetBstMidEndSepPunct{\mcitedefaultmidpunct}
{\mcitedefaultendpunct}{\mcitedefaultseppunct}\relax
\EndOfBibitem
\bibitem[Gupta and Patey(2014)Gupta, and Patey]{gupta2014structure}
Gupta,~R.; Patey,~G. \emph{J. Chem. Phys.} \textbf{2014}, \emph{141},
  064502\relax
\mciteBstWouldAddEndPuncttrue
\mciteSetBstMidEndSepPunct{\mcitedefaultmidpunct}
{\mcitedefaultendpunct}{\mcitedefaultseppunct}\relax
\EndOfBibitem
\bibitem[Limmer \latin{et~al.}(2013)Limmer, Merlet, Salanne, Chandler, Madden,
  Van~Roij, and Rotenberg]{Limmer:PRL:2013}
Limmer,~D.~T.; Merlet,~C.; Salanne,~M.; Chandler,~D.; Madden,~P.~A.;
  Van~Roij,~R.; Rotenberg,~B. \emph{Phys. Rev. Lett.} \textbf{2013},
  \emph{111}, 106102\relax
\mciteBstWouldAddEndPuncttrue
\mciteSetBstMidEndSepPunct{\mcitedefaultmidpunct}
{\mcitedefaultendpunct}{\mcitedefaultseppunct}\relax
\EndOfBibitem
\bibitem[Elmatad \latin{et~al.}(2010)Elmatad, Jack, Chandler, and
  Garrahan]{elmatad2010finite}
Elmatad,~Y.~S.; Jack,~R.~L.; Chandler,~D.; Garrahan,~J.~P. \emph{Proc. Natl.
  Acad. Sci. U.S.A.} \textbf{2010}, \emph{107}, 12793--12798\relax
\mciteBstWouldAddEndPuncttrue
\mciteSetBstMidEndSepPunct{\mcitedefaultmidpunct}
{\mcitedefaultendpunct}{\mcitedefaultseppunct}\relax
\EndOfBibitem
\bibitem[Mobley \latin{et~al.}(2006)Mobley, Chodera, and Dill]{mobley2006use}
Mobley,~D.~L.; Chodera,~J.~D.; Dill,~K.~A. \emph{J. Chem. Phys.} \textbf{2006},
  \emph{125}, 084902\relax
\mciteBstWouldAddEndPuncttrue
\mciteSetBstMidEndSepPunct{\mcitedefaultmidpunct}
{\mcitedefaultendpunct}{\mcitedefaultseppunct}\relax
\EndOfBibitem
\bibitem[Chodera \latin{et~al.}(2011)Chodera, Mobley, Shirts, Dixon, Branson,
  and Pande]{chodera2011alchemical}
Chodera,~J.~D.; Mobley,~D.~L.; Shirts,~M.~R.; Dixon,~R.~W.; Branson,~K.;
  Pande,~V.~S. \emph{Curr. Opin. Struc. Biol.} \textbf{2011}, \emph{21},
  150--160\relax
\mciteBstWouldAddEndPuncttrue
\mciteSetBstMidEndSepPunct{\mcitedefaultmidpunct}
{\mcitedefaultendpunct}{\mcitedefaultseppunct}\relax
\EndOfBibitem
\bibitem[Liu \latin{et~al.}(2013)Liu, Crocker, and Sinno]{sinno3}
Liu,~X.; Crocker,~J.~C.; Sinno,~T. \emph{J. Chem. Phys.} \textbf{2013},
  \emph{138}, 244111\relax
\mciteBstWouldAddEndPuncttrue
\mciteSetBstMidEndSepPunct{\mcitedefaultmidpunct}
{\mcitedefaultendpunct}{\mcitedefaultseppunct}\relax
\EndOfBibitem
\end{mcitethebibliography}

\begin{mcitethebibliography}{4}
\providecommand*\natexlab[1]{#1}
\providecommand*\mciteSetBstSublistMode[1]{}
\providecommand*\mciteSetBstMaxWidthForm[2]{}
\providecommand*\mciteBstWouldAddEndPuncttrue
  {\def\EndOfBibitem{\unskip.}}
\providecommand*\mciteBstWouldAddEndPunctfalse
  {\let\EndOfBibitem\relax}
\providecommand*\mciteSetBstMidEndSepPunct[3]{}
\providecommand*\mciteSetBstSublistLabelBeginEnd[3]{}
\providecommand*\EndOfBibitem{}
\mciteSetBstSublistMode{f}
\mciteSetBstMaxWidthForm{subitem}{(\alph{mcitesubitemcount})}
\mciteSetBstSublistLabelBeginEnd
  {\mcitemaxwidthsubitemform\space}
  {\relax}
  {\relax}

\bibitem[Patel \latin{et~al.}(2010)Patel, Varilly, and
  Chandler]{Patel:JPCB:2010}
Patel,~A.~J.; Varilly,~P.; Chandler,~D. \emph{J. Phys. Chem. B} \textbf{2010},
  \emph{114}, 1632 -- 1637\relax
\mciteBstWouldAddEndPuncttrue
\mciteSetBstMidEndSepPunct{\mcitedefaultmidpunct}
{\mcitedefaultendpunct}{\mcitedefaultseppunct}\relax
\EndOfBibitem
\bibitem[Patel \latin{et~al.}(2011)Patel, Varilly, Chandler, and
  Garde]{Patel:JSP:2011}
Patel,~A.~J.; Varilly,~P.; Chandler,~D.; Garde,~S. \emph{J. Stat. Phys.}
  \textbf{2011}, \emph{145}, 265 -- 275\relax
\mciteBstWouldAddEndPuncttrue
\mciteSetBstMidEndSepPunct{\mcitedefaultmidpunct}
{\mcitedefaultendpunct}{\mcitedefaultseppunct}\relax
\EndOfBibitem
\bibitem[K{\"a}stner and Thiel(2005)K{\"a}stner, and Thiel]{Kastner:JCP:2005}
K{\"a}stner,~J.; Thiel,~W. \emph{J. Chem. Phys.} \textbf{2005}, \emph{123},
  144104\relax
\mciteBstWouldAddEndPuncttrue
\mciteSetBstMidEndSepPunct{\mcitedefaultmidpunct}
{\mcitedefaultendpunct}{\mcitedefaultseppunct}\relax
\EndOfBibitem
\end{mcitethebibliography}
%
\providecommand{\latin}[1]{#1}
\providecommand*\mcitethebibliography{\thebibliography}
\csname @ifundefined\endcsname{endmcitethebibliography}
  {\let\endmcitethebibliography\endthebibliography}{}


\clearpage
\newpage

\setcounter{figure}{0}
\setcounter{section}{0}
\renewcommand{\thefigure}{S\arabic{figure}}
\renewcommand{\theequation}{S\arabic{equation}}

\begin{center}
\section{Supporting Information for \\``Sparse Sampling of Water Density Fluctuations in Interfacial Environments''}
\author{Erte Xi,}
\author{Richard C. Remsing,}
\author{Amish J. Patel*}\\
\textit{Department of Chemical and Biomolecular Engineering, University of Pennsylvania, Philadelphia, PA 19104, USA}\\
E-mail: amish.patel@seas.upenn.edu
\end{center}
\subsection{Derivation of the Umbrella Sampling Equation}
Here we derive the umbrella sampling equation used in the main text, for a 
a biased Hailtonian, $\mathcal{H}_{\phi} = \mathcal{H}_0 + U_{\phi} (\tilde N_v)$.
The probability, $P_v(\tilde N)$, of observing $\tilde N$ coarse-grained waters in the unbiased ensemble is given by:
\begin{equation}
P_v(\tilde N) = \langle \delta(\tilde N_v- \tilde N) \rangle _0 = \frac{1}{Q_0} \int \delta(\tilde N_v- \tilde N) e^{-\beta \mathcal{H}_0}
\label{eq:pvn-def}
\end{equation}
where $Q_0\equiv\int e^{-\beta \mathcal{H}_0}$ is the partition function of the unbiased ensemble.
Employing a straightforward reweighting strategy, we get
\begin{equation}
P_v(\tilde N) = \left(\frac{Q_{\phi}}{Q_0}\right)\frac{1}{Q_{\phi}} \int \delta(\tilde N_v- \tilde N) e^{-\beta \mathcal{H}_{\phi}} e^{\beta U_{\phi}({\tilde N}_v)}
\label{eq:pvn-rew}
\end{equation}
where $Q_{\phi}\equiv\int e^{-\beta \mathcal{H}_{\phi}}$ is the partition function of the biased ensemble.
Finally, recognizing that the delta function allows us to pull the $e^{\beta U_{\phi}({\tilde N}_v)}$ term outside the integral, we get
\begin{equation}
P_v(\tilde N) =   \left(\frac{Q_{\phi}}{Q_0}\right)  e^{\beta U_{\phi}(\tilde N)}  \langle \deltaNt \rangle_{\phi}
\label{eq:pvn-final}
\end{equation}
Taking the logarithm of both sides and recognizing that $\langle \delta(\tilde N_v- \tilde N) \rangle_{\phi}$ is simply the probability, $P_v^{\phi}(\tilde N)$, of observing $\tilde N$ coarse-grained waters in $v$ in the biased ensemble, we get the central result of umbrella sampling as
\begin{equation}
-\ln P_v(\tilde N) = - \ln P_v^{\phi} (\tilde N) -\beta U_{\phi} (\tilde N) -\ln \left(\frac{Q_{\phi}}{Q_0}\right)
\label{eq:pvn-ln}
\end{equation}
%

\subsection{Getting $P_v(N)$ using the sparse sampling method}
\begin{figure}
\begin{center}
\includegraphics[width=0.9\textwidth]{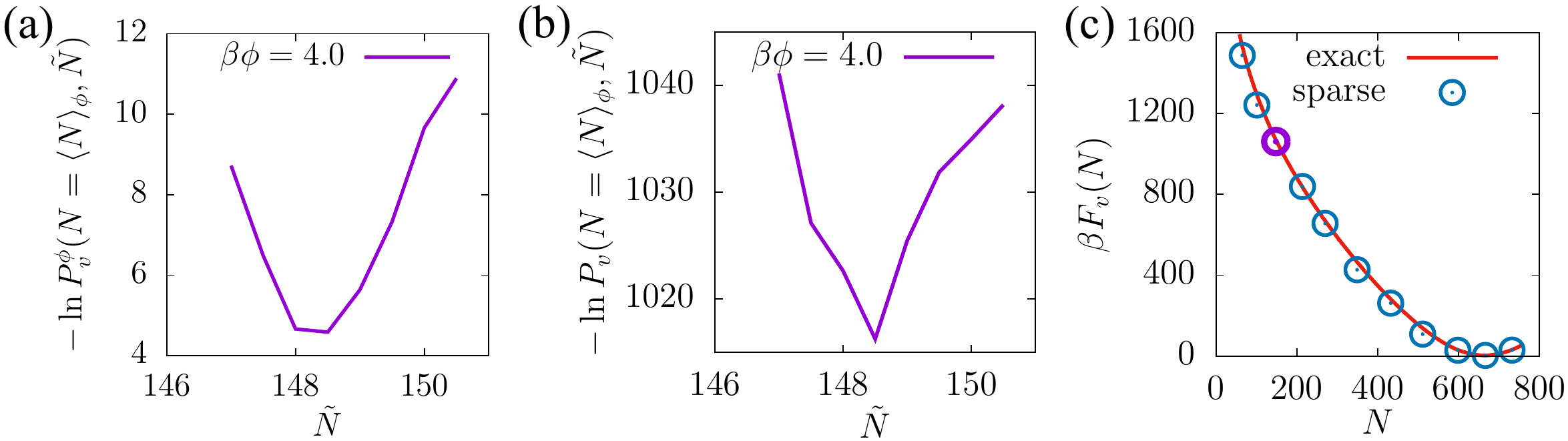}
\caption{
Obtaining $P_v(N)$ for ubiquitin hydration shell.
(a) The $\beta \phi=4$ biased simulation was used to estimate the biased joint distribution, $P_v(N,\tilde N)$, at $N = \avgN_\phi$.
(b) The corresponding unbiased joint distribution was obtained using equation~\ref{eq:jpvn-ln}.
(c) Each biased simulation was used to sparsely sample $F_v(N)$ by integrating the respective unbiased joint distributions using equation~\ref{eq:jpvn-int}; the integral of the joint distribution from the $\beta \phi=4.0$ ensemble is shown in purple. The resulting $F_v(N)$ agrees well with the exact results obtained using the Indirect Umbrella Sampling method~\cite{Patel:JPCB:2010,Patel:JSP:2011}.
}
\label{fig:pvn}
\end{center}
\end{figure}

%
In the main text, we develop a sparse sampling method to estimate $P_v(\tilde N)$, the probability of observing $\tilde N$ coarse-grained waters in a volume, $v$, of interest.  
Here we show that using the same framework, a closely related discrete order parameter, such as the number of waters in $v$, can also be sparse sampled, that is, the probability distribution, $P_v(N)$, of observing $N$ waters in $v$ can also be estimated. 
As in the previous section, the joint probability distribution, $P_v(N,\tilde N)$, of observing $N$ waters and $\tilde N$ corse-grained waters in the unbiased ensemble is given by:
\begin{equation}
-\ln P_v(N, \tilde N) = - \ln P_v^{\phi} (N, \tilde N) -\beta U_{\phi} (\tilde N) -\ln \left(\frac{Q_{\phi}}{Q_0}\right)
\label{eq:jpvn-ln}
\end{equation}
The central idea is to bias the system using the coarse-grained order parameter, $\tilde N$, but collect both $\tilde N$ and $N$ data from the biased simulations to obtain $P_v^{\phi} (N, \tilde N)$, the biased joint distribution, which is the first term on the right side of Equation~\ref{eq:jpvn-ln}.
The biasing potential (the second term) is known analytically and the free energy differences (the third term)  between biased and unbiased ensemble can be readily estimated using thermodynamic integration, as discussed in the main text. 
Finally, $\pvn$ can be estimated by an integrating the unbiased joint distribution over $\tilde N$, 
\begin{equation}
P_v(N) = \int P_v(N, \tilde N) d\tilde N
\label{eq:jpvn-int}
\end{equation}
To illustrate Equation~\ref{eq:jpvn-ln} and Equation~\ref{eq:jpvn-int}, we apply this strategy to the hydration shell of the protein, ubiquitin, which was discussed in the main text. 
An example of the biased and corresponding unbiased joint distributions at an $N$-value (in this case, $N = \avgN_\phi$ with $\beta\phi=4$) of interest, are shown in Figures~\ref{fig:pvn}a and~\ref{fig:pvn}b respectively. 
Figure~\ref{fig:pvn}c shows that the sparse sampled free energies, $\beta F_v(N) \equiv -\ln P_v(N)$, obtained from Equation~\ref{eq:jpvn-int}, agree well with the exact result computed by the Indirect Umbrella Sampling method~\cite{Patel:JPCB:2010,Patel:JSP:2011}. 
%

\subsection{Comparison between the Sparse Sampling Method and Umbrella Integration}
%
The Umbrella Integration (UI) method, proposed by K{\"a}stner and Thiel~\cite{Kastner:JCP:2005}, utilizes principles of both umbrella sampling and thermodynamic integration to estimate free energies, akin to the sparse sampling method that we present here.
However, there are important distinctions between the two methods, which we discuss below.
\\1) While we explicitly estimate
the free energy differences between the biased and unbiased ensembles
using thermodynamic integration, UI prescribes eliminating them from the analysis altogether by differentiating Equation
2 of the main text.
$\partial F_v/\partial\Nt$-values are then estimated using biased simulations, and integrated numerically to yield estimates of $F_v(\Nt)$. 
\\2) The authors focus on the widely-used harmonic form of the biasing potential, and show that in the limiting case of a stiff spring, UI becomes equivalent to thermodynamic integration.
In contrast, we use a linear biasing potential, which contributes significantly to the efficiency of our method; in particular, the linear potential ensures that $\avgNt_\phi$ decreases monotonically with $\phi$ (because $\partial\avgNt_\phi/\partial\phi=-\beta\varNt_\phi<0$), and enables estimation of $F_\phi$ using only a few biased simulations.
\\3) Finally, as recognized by its authors, UI does not, in principle, require overlap between adjacent windows.
In practice, however, the functional form of $\partial F_v/\partial\Nt$ is not known a priori, so accurate estimates of $F_v(\Nt)$ rely on estimates of $\partial F_v/\partial\Nt$ at finely spaced $\Nt$-values.
Indeed, the authors use UI not to perform sparse sampling, but to illustrate that it leads to smaller statistical errors relative to WHAM (Weighted Histogram Analysis Method).
%

\section{Fluctuations in the Hydration Shell of the Protein, Hydrophobin}
\begin{figure}
\begin{center}
\includegraphics[width=0.6\textwidth]{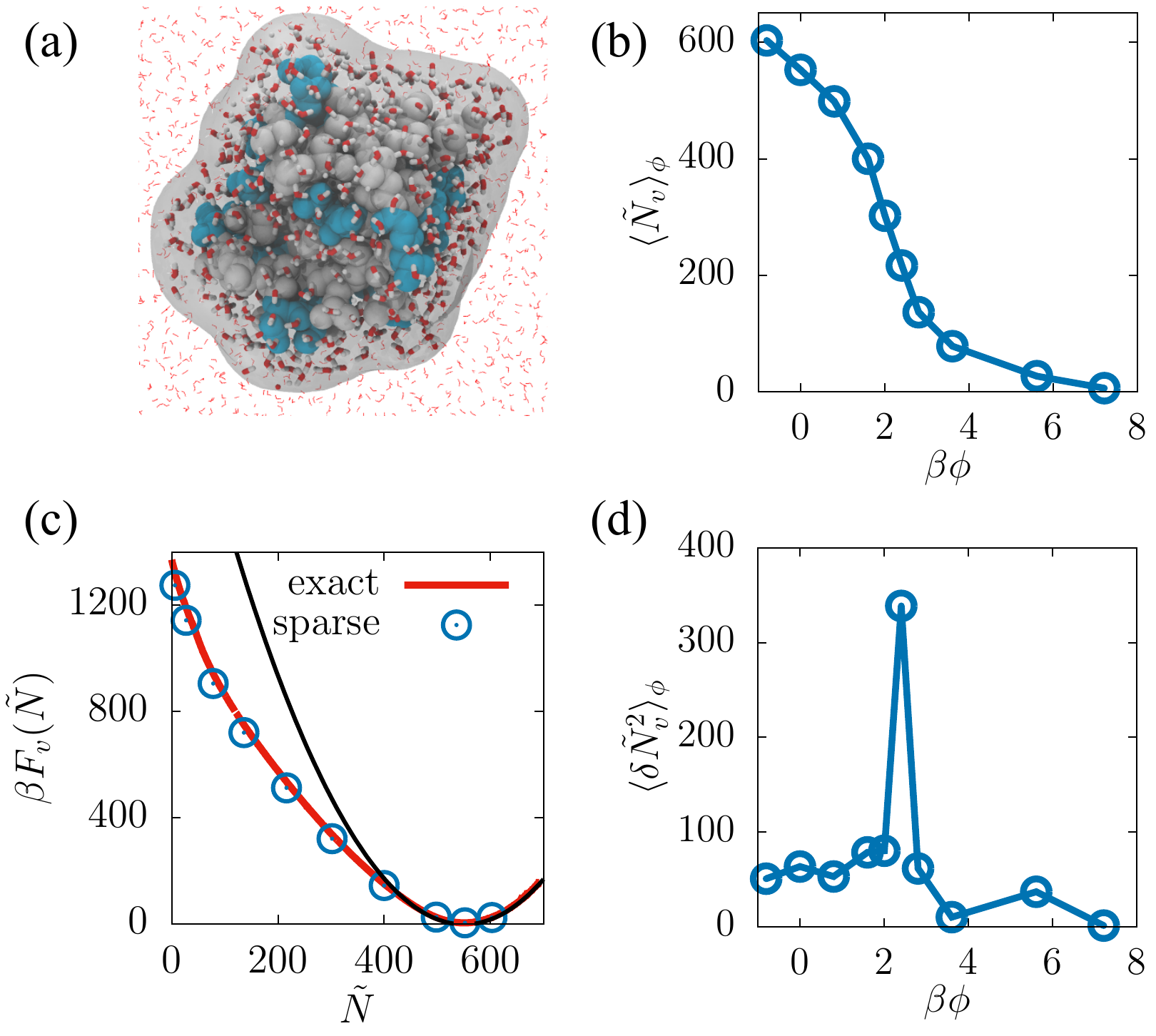}
\caption{
Applying the sparse sampling method to characterize the free energetics of water density fluctuations in the hydration shell of the protein, hydrophobin II. 
(a) Simulation snapshot of the system, where the hydrophobin (blue/white space-fill representation) is solvated in bulk water (red/white lines). The hydration shell (gray) contains roughly 600 water molecules (sticks) on average. 
(b) The response, $\avgNt_\phi$, to a biasing potential of strength $\phi$, displays sigmoidal behavior. 
(c) The sparsely sampled $F_v(\tilde N)$ agrees well with the exact result obtained using umbrella sampling, and displays a pronounced low-$\tilde N$ fat tail. 
(d) The susceptibility displays a peak around $\beta \phi = 2$, suggestive of a collective dewetting transition in the hydration shell of the protein. 
}
\label{fig:hfb}
\end{center}
\end{figure}

Here, we use the sparse sampling method to estimate $P_v(\tilde N)$ in the hydration shell of the protein, hydrophobin II (PDB ID: 2B97), which is known to have a large hydrophobic patch on the protein surface.
As shown in Figure~\ref{fig:hfb}, the protein displays characteristics that are quite similar to that of the protein, ubiquitin, discussed in the main text.  
%

\providecommand{\latin}[1]{#1}
\providecommand*\mcitethebibliography{\thebibliography}
\csname @ifundefined\endcsname{endmcitethebibliography}
  {\let\endmcitethebibliography\endthebibliography}{}

\end{document}